\documentclass[twoside,twocolumn,9pt]{article}
\usepackage{extsizes}
\usepackage[super,sort&compress,comma]{natbib} 
\usepackage[version=3]{mhchem}
\usepackage[left=1.5cm, right=1.5cm, top=1.785cm, bottom=2.0cm]{geometry}
\usepackage{balance}
\usepackage{times,mathptmx}
\usepackage{sectsty}
\usepackage{graphicx} 
\usepackage{lastpage}
\usepackage[format=plain,justification=justified,singlelinecheck=false,font={stretch=1.125,small,sf},labelfont=bf,labelsep=space]{caption}
\usepackage{float}
\usepackage{fancyhdr}
\usepackage{fnpos}
\usepackage[english]{babel}
\addto{\captionsenglish}{%
  
}
\usepackage{array}
\usepackage{siunitx}
\usepackage{droidsans}
\usepackage{charter}
\usepackage[T1]{fontenc}
\usepackage[usenames,dvipsnames]{xcolor}
\usepackage{setspace}
\usepackage[compact]{titlesec}
\usepackage{hyperref}
\usepackage{amsfonts} 
\usepackage{amsmath} 
\usepackage{amssymb} 
\usepackage{bm}
\usepackage{mathtools}

\usepackage{epstopdf}

\definecolor{cream}{RGB}{222,217,201}

\newcommand\figref[2]{Fig.~\ref{#1}#2}

\newcommand{\D}{\mathrm{d}}

\newcommand\op[3]{\bm{#1}^{#2}_{#3}}

\newcommand\R[1]{\op{r}{}{#1}}

\newcommand{\sy}{\sigma_{\rm y}}

\newcommand{\gd}{\dot{\gamma}}

\newcommand{\gps}{\dot{\gamma}_{\rm ps}}

\newcommand{\DUinf}{\bm{\nabla}\bm{U}^{\infty}}
\newcommand{\Einf}{\bm{E}^{\infty}}
\newcommand{\Ominf}{\bm{\Omega}^{\infty}}

\newcommand{\gSigma}{\bm{\Sigma}}

\newcommand{\gF}{\bm{F}}

\newcommand\gdot{\bm{\cdot}}

\newcommand{\er}{\bm{e}_r}

\newcommand{\eroer}{\er\otimes\er}

\def\identity{\;\mbox{l\hspace{-0.55em}1}}

\renewcommand{\eqref}[1]{Eqn~(\ref{#1})}

\begin{document}

\pagestyle{fancy}
\thispagestyle{plain}
\fancypagestyle{plain}{

\renewcommand{\headrulewidth}{0pt}
}

\makeFNbottom
\makeatletter
\renewcommand\LARGE{\@setfontsize\LARGE{15pt}{17}}
\renewcommand\Large{\@setfontsize\Large{12pt}{14}}
\renewcommand\large{\@setfontsize\large{10pt}{12}}
\renewcommand\footnotesize{\@setfontsize\footnotesize{7pt}{10}}
\makeatother

\renewcommand{\thefootnote}{\fnsymbol{footnote}}
\renewcommand\footnoterule{\vspace*{1pt}%
\color{cream}\hrule width 3.5in height 0.4pt \color{black}\vspace*{5pt}} 
\setcounter{secnumdepth}{5}

\makeatletter 
\renewcommand\@biblabel[1]{#1}            
\renewcommand\@makefntext[1]%
{\noindent\makebox[0pt][r]{\@thefnmark\,}#1}
\makeatother 
\renewcommand{\figurename}{\small{Fig.}~}
\sectionfont{\sffamily\Large}
\subsectionfont{\normalsize}
\subsubsectionfont{\bf}
\setstretch{1.125} 
\setlength{\skip\footins}{0.8cm}
\setlength{\footnotesep}{0.25cm}
\setlength{\jot}{10pt}
\titlespacing*{\section}{0pt}{4pt}{4pt}
\titlespacing*{\subsection}{0pt}{15pt}{1pt}

\fancyfoot{}
\fancyfoot[LO,RE]{\vspace{-7.1pt}\includegraphics[height=9pt]{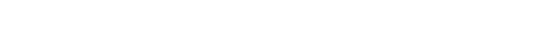}}
\fancyfoot[CO]{\vspace{-7.1pt}\hspace{13.2cm}\includegraphics{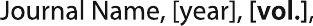}}
\fancyfoot[CE]{\vspace{-7.2pt}\hspace{-14.2cm}\includegraphics{head_foot/RF}}
\fancyfoot[RO]{\footnotesize{\sffamily{1--\pageref{LastPage} ~\textbar  \hspace{2pt}\thepage}}}
\fancyfoot[LE]{\footnotesize{\sffamily{\thepage~\textbar\hspace{3.45cm} 1--\pageref{LastPage}}}}
\fancyhead{}
\renewcommand{\headrulewidth}{0pt} 
\renewcommand{\footrulewidth}{0pt}
\setlength{\arrayrulewidth}{1pt}
\setlength{\columnsep}{6.5mm}
\setlength\bibsep{1pt}

\makeatletter 
\newlength{\figrulesep} 
\setlength{\figrulesep}{0.5\textfloatsep} 

\newcommand{\topfigrule}{\vspace*{-1pt}%
\noindent{\color{cream}\rule[-\figrulesep]{\columnwidth}{1.5pt}} }

\newcommand{\botfigrule}{\vspace*{-2pt}%
\noindent{\color{cream}\rule[\figrulesep]{\columnwidth}{1.5pt}} }

\newcommand{\dblfigrule}{\vspace*{-1pt}%
\noindent{\color{cream}\rule[-\figrulesep]{\textwidth}{1.5pt}} }

\makeatother

\twocolumn[
  \begin{@twocolumnfalse}
{\includegraphics[height=30pt]{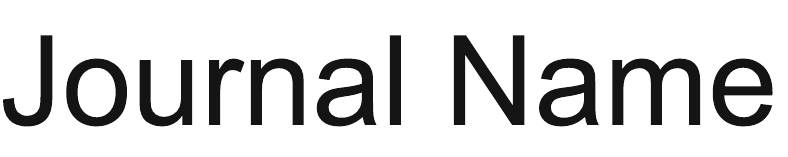}\hfill\raisebox{0pt}[0pt][0pt]{\includegraphics[height=55pt]{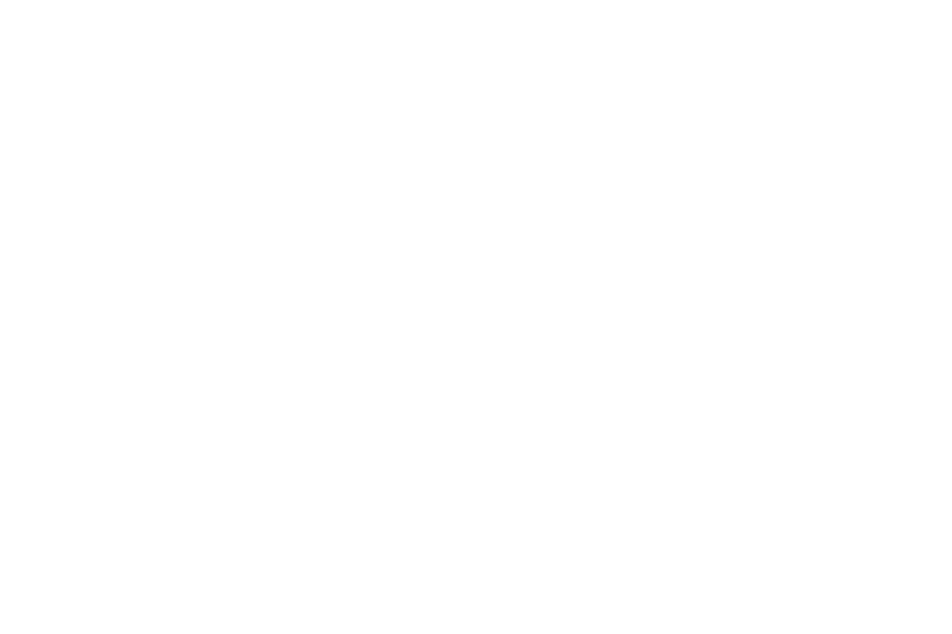}}\\[1ex]
\includegraphics[width=18.5cm]{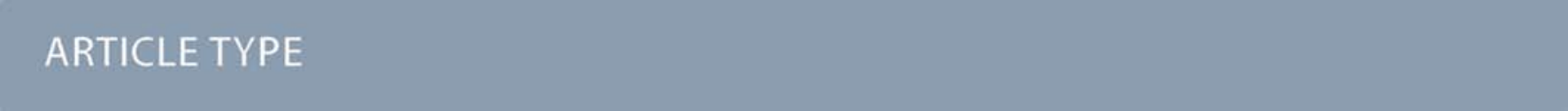}}\par
\vspace{1em}
\sffamily
\begin{tabular}{m{4.5cm} p{13.5cm} }

\includegraphics{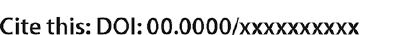} & \noindent\LARGE{\textbf{%
Dynamics of microstructure anisotropy and rheology of soft jammed suspensions
}} \\
\vspace{0.3cm} & \vspace{0.3cm} \\

 & \noindent\large{Nicolas Cuny,\textit{$^{a}$} Eric Bertin,\textit{$^{a}$} and Romain Mari$^{\ast}$\textit{$^{a}$}} \\

\includegraphics{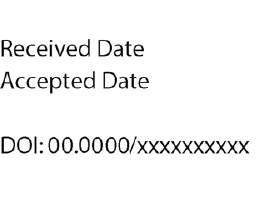} & \noindent\normalsize{We explore the rheology predicted by a recently proposed constitutive model for jammed suspensions of soft elastic particles derived from particle-level dynamics [Cuny \textit{et al.}, Phys. Rev. Lett. \textbf{127}, 218003 (2021)]. Our model predicts that the orientation of the anisotropy of the microstructure, governed by an interplay between advection and contact elasticity, plays a key role at yielding and in flow. 
It generates normal stress differences contributing significantly to the yield criterion and Trouton ratio.
It gives rise to non-trivial transients such as stress overshoots in step increases of shear rate, residual stresses after flow cessation and power-law decay of the shear rate in creep. 
Finally, it explains the collapse of storage modulus as measured in parallel superposition for a yielded suspension.} \\

\end{tabular}

 \end{@twocolumnfalse} \vspace{0.6cm}

  ]

\renewcommand*\rmdefault{bch}\normalfont\upshape
\rmfamily
\section*{}
\vspace{-1cm}


\footnotetext{\textit{$^{a}$~Univ. Grenoble Alpes, CNRS, LIPhy, 38000 Grenoble, France, E-MAIL: romain.mari@univ-grenoble-alpes.fr}}




\section{Introduction}

Soft jammed suspensions are a class of materials made of soft, elastic particles immersed in a fluid, with a concentration large enough to create a continuous elastic network of contacts, which induces a yield stress.
Concentrated suspensions of microgels, concentrated emulsions or wet foams belong to this family~\cite{bonnecaze_micromechanics_2010}.
Despite being common occurrence in industrial contexts, in particular in the food and cosmetic industries, the rheology of these materials is still being explored, 
and many open questions remain, not only regarding quantitative characterization, but also regarding what observables are most suited to describe these systems, especially in their yielded phase~\cite{coussotYieldStressFluid2014}.
The yield point and steady rheology under flow have been the main focus of the literature, 
and in most cases results concern simple shear, for which only the shear stress is reported. 

Based on simple shear results, a qualitative rheological picture for homogeneous flows (which are not always the norm~\cite{divoux_transient_2010,divoux_stress_2011}) has emerged. 
Soft jammed suspensions follow a Herschel-Bulkley rheology with the shear stress $\sigma$ related to the shear rate $\dot\gamma$ as $\sigma = \sigma_\mathrm{y,s}+ k \dot\gamma^n$, with $\sigma_\mathrm{y,s}$ the (simple shear) yield stress, $k>0$ the consistency parameter and the flow parameter $n \approx 0.5$~\cite{piau_carbopol_2007,moller_attempt_2009,divoux_transient_2010,seth_micromechanical_2011,liu_universality_2018}.
Before this steady-state stress is reached, soft jammed suspensions sometimes exhibit a stress overshoot, that is, a transient maximum before decreasing to the steady-state stress value \cite{partal_transient_1999,batista_colored_2006,coussot_macroscopic_2009,divoux_stress_2011,dinkgreve_different_2016,younes_elusive_2020, khabazTransientDynamicsSoft2021}. 
In contrast, tensorial and temporal aspects only quite recently received some attention~\cite{de_cagny_yield_2019,divoux_stress_2011,mohan_microscopic_2013,ngouambaElastoplasticBehaviorYield2019}, 
despite the importance of the question in virtually any practical application, which involve unsteady conditions and varying flow geometries.

The history of constitutive modelling is consequently equally steady-simple-shear centric. 
Many phenomenological models are designed around the scalar HB rheology, from which a tensorial extension is postulated~\cite{oldroydRationalFormulationEquations1947,saramito_new_2007,saramito_new_2009,balmforth_yielding_2014}.
These models involve a von Mises criterion for the yield, based on the second invariant of the deviatoric stress tensor $\bm{\Sigma}'$, stating that yield occurs when $\sqrt{\bm{\Sigma}':\bm{\Sigma}'/2} = \tau_\mathrm{c}$, with $\tau_\mathrm{c}$ a material property (and therefore independent of the type of deformation under which yield is achieved).
The validity of the von Mises yield criterion, as well as the experimental methodologies to test it, have been recently actively debated~\cite{ovarlezThreedimensionalJammingFlows2010,germanFormationViscoplasticDrops2010,shaukatShearMediatedElongational2012,martinieApparentElongationalYield2013,zhangYieldingFlowSoftJammed2018,thompsonYieldStressTensor2018,de_cagny_yield_2019,thompsonRheologicalMaterialFunctions2020}.
In particular, there is no consensus as to whether measuring the yield shear stress in a simple shear setup is enough to determine $\tau_\mathrm{c}$ if the von Mises criterion holds, or equivalently, whether normal stresses at yield under simple shear are negligible in front of the shear stress~\cite{thompsonYieldStressTensor2018,de_cagny_yield_2019}.

Separately, dynamical aspects of soft jammed suspensions are also poorly understood.
Shear under constant imposed stress set at the yield value induces a creep behavior, with a power law decay of the shear rate as a function of time~\cite{divoux_transient_2010,lidon_power-law_2017}. 
Flow cessation leads to residual stresses, which amplitude decreases with the stress or rate applied in pre-shear~\cite{mohan_microscopic_2013,mohan_build-up_2014,vasishtResidualStressAthermal2021}.
Finally, while visco-elasticity in unsheared materials is well studied, the linear response under flow (often called parallel superposition) shows interesting unexplained features: at fixed frequency the storage modulus strongly decreases as a function of applied stress above yield, 
while the loss modulus shows a more moderate decay, with a possible upward jump at yielding~\cite{benmouffok-benbelkacemNonlinearViscoelasticityTemporal2010,ngouambaElastoplasticBehaviorYield2019}, suggesting a loss of elastic integrity of the particle contact network which is difficult to reconcile with the picture of densely packed particles above their jamming point.

Using a recent constitutive model derived from particle-level dynamics~\cite{CunyPRL21}, we here address tensorial and dynamical aspects of the rheology of soft jammed suspensions, 
namely the contribution of normal stresses to the von Mises criterion and Trouton ratio under flow, the linear response under shear, and the transients under step changes in applied stress or rate.
Thanks to the microscopic grounding of our model, we can relate the rheology to microstructure evolution.
We will see that in many cases, the tensorial nature of the anisotropy of the elastic contact network, which couples to the vorticity when straining is present, plays a simple but central role in apparently complex rheological responses.
The tensorial elasticity/advection interplay is particularly simple in slow flows, that is, those for which the deformation rate is small in front of the inverse elastic timescale. 
For these, the total stress is essentially coming from the elastic particle stress.
We show that the amplitude of the deviatoric stress relaxes elastically, and thus has a much faster dynamics than the orientation of its principal axes, which is advected.
We find that this amplitude/orientation timescale decoupling is leading to possibly deceiving transient behaviors for the shear component of the stress in simple shear. More generally, for flows with both finite strain and finite vorticity, one needs to acknowledge the presence of normal stress differences to build a coherent physical picture.

\section{Constitutive model}

\subsection{Stress evolution}

In a recent work~\cite{CunyPRL21,CunyPRE21}, we derived from microscopic dynamics a constitutive law for a two-dimensional system of overdamped frictionless harmonic disks above the jamming transition. This system is an idealized model for soft jammed suspensions such as concentrated emulsions or microgels~\cite{durian_foam_1995,ikedaUnifiedStudyGlass2012,vagbergPressureDistributionCritical2014,liu_universality_2018}. 
In this model, disks of radius $a$ interact through radial contact repulsion forces deriving from the harmonic potential $V(r) = af_0 (1-r/2a)^2\Theta(1 - r/2a)$ with $\Theta$ the Heaviside function.
They are further subject to a viscous drag $-\lambda_{\rm f} (\dot{\bm{r}}_\mathrm{i} - \bm{u}^{\infty}(\bm{r}_\mathrm{i}))$, where $\dot{\bm{r}}_\mathrm{i}$ is the velocity of particle $i$, and $\bm{u}^{\infty}(\bm{r}_\mathrm{i})$ the velocity at the particle position of a fictitious fluid  that induces shear in the system.
This model is characterized by an elastic timescale $\tau_0 \equiv f_0/(\lambda_{\rm f} a)$.

Under a uniform applied flow, characterized by its velocity gradient $\nabla \bm{u}^\infty$ (which we define as $(\nabla \bm{u}^\infty)_{ij} = \partial_j u^\infty_i$), there is a regime, away from jamming and yielding criticalities, for which flow is homogeneous in space and time. This stands in contrast to the plastic-event dominated regime close to the yielding transition~\cite{nicolas_deformation_2018}, and the soft-mode dominated regime close to jamming~\cite{lernerUnifiedFrameworkNonBrownian2012}.
In Refs.~\cite{CunyPRL21,CunyPRE21}, we derived in this regime the following evolution equation for the deviatoric part of the particle (or elastic) stress tensor $\bm{\Sigma}'$ 
\begin{equation}
  \label{eq:Sigmaprim}
  \dot{\bm{\Sigma}}' = \kappa(\phi)\Einf + \Ominf \gdot \gSigma' - \gSigma' \gdot \Ominf + \left[\beta(\phi) - \xi(\phi) \left(\gSigma':\gSigma'\right)\right]\gSigma'\, ,
\end{equation}
where the strain-rate tensor $\Einf$ and the vorticity tensor $\Ominf$ are respectively the symmetric and antisymmetric parts of $\nabla \bm{u}^\infty$.
The coefficients $\kappa$, $\beta$ and $\xi$ have known expressions as a function of the area fraction $\phi$ and of particle-level parameters. The key steps of the derivation are recalled in Appendix~A.
In short, the basic idea is to derive an evolution equation for the stress tensor from the evolution equation of the pair correlation function 
(the athermal equivalent of the Smoluchovski equation~\cite{russelRheologySuspensionsCharged1978,felderhofEffectBrownianMotion1983,bradyRheologicalBehaviorConcentrated1993,nazockdast_microstructural_2012}), using the virial definition of the stress to relate microstructure to stress. 
A set of physically plausible approximations then allows one to get the closed evolution equation Eq.~(\ref{eq:Sigmaprim}) for the deviatoric part $\bm{\Sigma}'$ of the particle stress tensor.
This stress evolution focuses on the regime where the total stress is mostly coming from the elastic forces, rather than viscous dissipation, that is, for $\dot\gamma \equiv \sqrt{2\Einf:\Einf} \ll \tau_0^{-1}$ (small Weissenberg numbers).
We stress here that \eqref{eq:Sigmaprim} is for a two-dimensional system, for which the usual advection term proportional to $\Einf \gdot \gSigma' + \gSigma' \gdot \Einf$ has vanishing deviatoric contribution.

This constitutive model is formally quite similar to the Saramito model~\cite{saramito_new_2007} above yield, except that our microscopic derivation leaves no freedom to the parameters, which are entirely determined from microscopic properties.
Indeed, the parameters $\kappa$, $\beta$ and $\xi$ are known analytical functions (albeit complicated ones) of the area fraction $\phi$ of the suspension.
Note that dimensional analysis imposes that they also are functions of the particle radius and stiffness and of the fluid viscosity as $\kappa \propto f_0/a = \lambda_\mathrm{f} \tau_0$, $\beta\propto \tau_0^{-1}$ and 
$\xi\propto a^2/(\tau_0 f_0^2)= 1/(\lambda_\mathrm{f}^2 \tau_0^3)$.
Linearizing these functions at the jamming transition located at $\phi_\mathrm{J}$, we find  $\kappa a/f_0\approx 1.19 - 0.099 \Delta \phi$,
$\beta\tau_0\approx 0.16 +0.76\Delta \phi$, and
$\xi \tau_0 f_0^2 /a^2\approx 0.62 +0.0054\Delta \phi$, with $\Delta \phi = \phi-\phi_{\rm J}$.
As the approximations involved in deriving \eqref{eq:Sigmaprim} 
assume that we are not very close to jamming, these expressions should  
be taken as approximations rather than controlled perturbative expansions around jamming.

In spite of the interest of having a microscopically grounded constitutive model,
a possible drawback is that the model has no fitting flexibility, and its quantitative predictions depend on the details of the quality of the closures used in the derivation.
For instance, it predicts a yield stress~\cite{CunyPRL21} for volume fractions larger than a jamming volume fraction $\phi_\mathrm{J} = 1.25$, which is around \SI{50}{\percent} larger than the actual measured values in simulations~\cite{durian_foam_1995}.
Finally, this model is for uniformly flowing suspensions. It does not possess an elastic branch before yield, nor even a static mechanical equilibrium inside the yield surface.
Furthermore, because it does not consider spatial and temporal fluctuations like plastic events, it predicts a Bingham rheology~\cite{CunyPRL21}, as opposed to a Herschel-Bulkley one with $n<1$.

Nonetheless, our model is the first one to explicitly associate a microscopic picture 
with the macroscopic rheology of concentrated suspensions of elastic particles. 
It thus provides a frame of thought to approach these systems, with clearly stated 
organizing principles, such as the microstructure-stress relation we present below,
which strengths and weaknesses can be discussed.

\begin{figure}
    \centering
  \includegraphics[width=\linewidth]{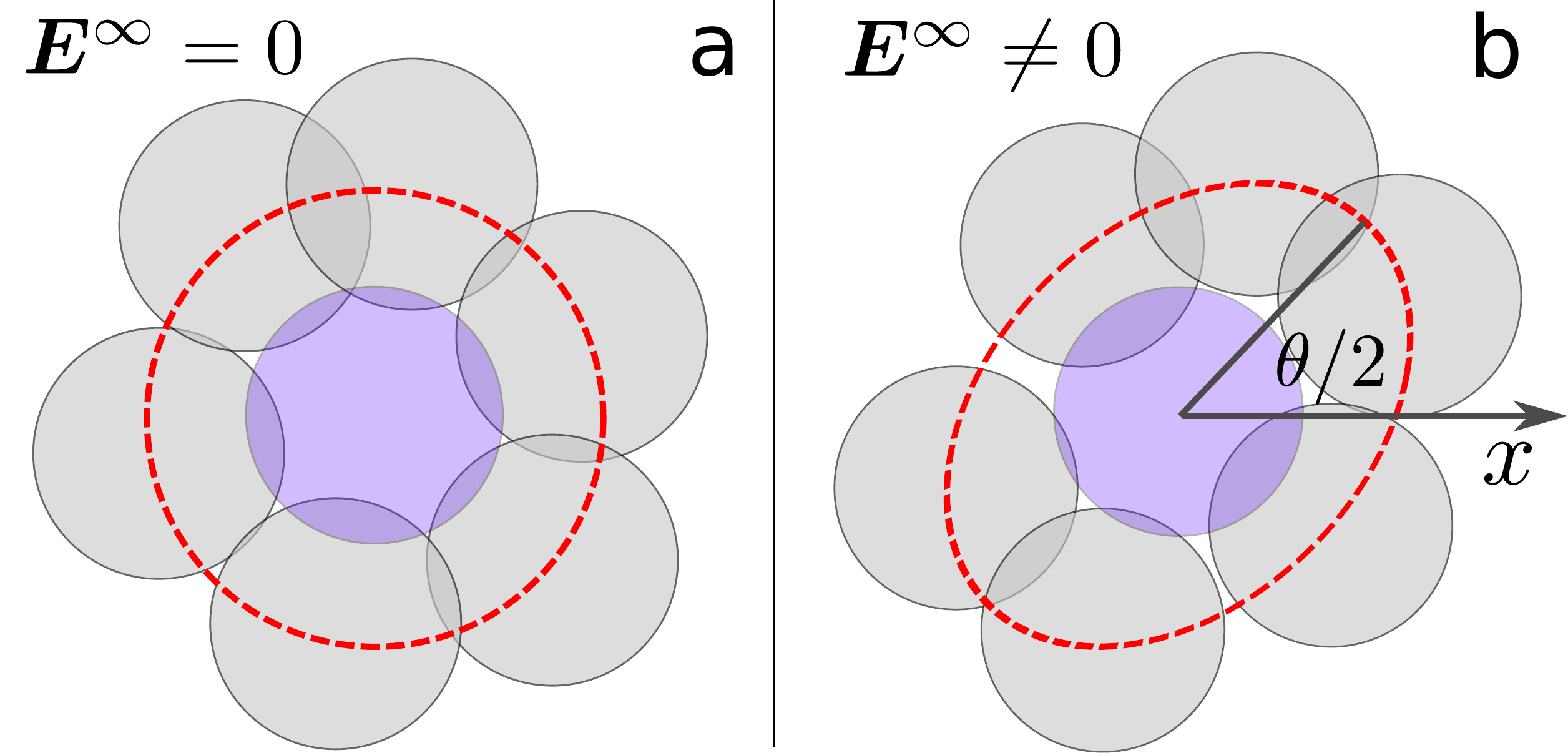}
  \caption{(a) A soft jammed suspension in isotropic conditions, without externally applied deformation. The central particle is surrounded by a shell of nearest neighbors which centers are located near the red dashed line. (b) Under an applied strain rate, the microstructure develops an anisotropy, here represented by the deformation of the initially circular nearest-neighbor peak location. The anisotropy is characterized by an amplitude $S$ (see main text for definition) and an orientation $\theta/2$ of its main axis with respect to the $x$ direction.}
  \label{fig:polar_visual}
\end{figure}

\subsection{Characterizing generic steady uniform flows}
In this article, we will study in depth tensorial aspects of our model, testing in particular its dependence on flow geometry. 
For this, it is useful to define the entire set of possible steady uniform flow geometries in two dimensions.
For a given steady uniform flow defined by its velocity gradient $\nabla \bm{u}^\infty$, we call flow geometry (or ``shape'') the tensor $\bm{K}^\infty = \nabla \bm{u}^\infty/\sqrt{\nabla \bm{u}^\infty:\nabla \bm{u}^\infty}$. 
(This definition of $\bm{K}^\infty$ is a special case of the more general, frame-indifferent definition used for arbitrary unsteady and non-uniform flows, that we give in Appendix B.)
While in two dimensions $\nabla \bm{u}^\infty$ has three independent values, the flow geometry, being the normalized $\nabla \bm{u}^\infty$, has only two.
Moreover, because all flow geometries differing only by a solid rotation are equivalent, we can without loss of generality restrict our exploration to a one-dimensional family parametrized by a single angle $\alpha \in [0,\pi/2]$,
\begin{equation}
\bm{K}^\infty = \frac{1}{\sqrt{2}}
                \begin{pmatrix}
                    \cos \alpha \cos 2\alpha &  2\sin\alpha + \sin \alpha \cos 2\alpha \\
                    \sin \alpha \cos 2\alpha &  -\cos \alpha \cos 2\alpha
    \end{pmatrix}\, .
    \label{eq:flow_family}
\end{equation}
Planar extensional flow corresponds to $\alpha = 0$, simple shear is obtained for $\alpha=\pi/4$, while for $\alpha=\pi/2$ the flow is purely rotational (and thus no yield is possible for this value).
The derivation of \eqref{eq:flow_family} is given in Appendix B.

\subsection{Microstructure-stress relation}

We will exploit the microscopic grounding of the model in order to relate rheological features 
to microstructural ones at the level of the pair correlation function $g(\bm{r})$ between particles. 
Indeed, in our model we find that the deviatoric stress $\bm{\Sigma}'$ 
is proportional to the fabric tensor $\bm{Q}$, $\bm{\Sigma}' = \mu(\phi)\bm{Q}$. This tensor encodes the lowest order in anisotropy of the pair correlation function~\cite{CunyPRE21}, and is defined as
\begin{equation}
    \bm{Q} = \frac{\phi^2}{2\pi^2a^4}\int_{|\bm{r}|\le 2a}\left[\bm{r}\otimes \bm{r} - \frac{|\bm{r}|^2}{2} \identity\right]g(\bm{r})\D\bm{r}\,.
\end{equation}
$\bm{Q}$ is traceless, so it has two eigenvalues equal in amplitude but of opposite signs.
The eigenvector with positive eigenvalue $\lambda$ (resp. negative eigenvalue $-\lambda$) 
corresponds to the direction where particle contacts are the least (resp. the most) compressed under flow, as depicted in Fig.~\ref{fig:polar_visual}{b}.
In this figure we defined as $\theta/2$ the angle that the eigenvector with positive eigenvalue makes with the $x$ direction.

Because $\bm{\Sigma}' = \mu(\phi) \bm{Q}$, $\theta$ also quantifies the tilt of the stress anisotropy, and $\bm{\Sigma}'$ takes the form
\begin{equation}
    \bm{\Sigma}' = S\begin{pmatrix}
                     \cos \theta & \sin \theta \\
                     \sin \theta & -\cos \theta
                    \end{pmatrix}\, ,
    \label{eq:polar_Sigma}
\end{equation}
where we define $S$ as the amplitude of the stress anisotropy.
The pair $(S, \theta)$ is the polar coordinate system associated with the deviatoric stress tensor, and is related to the usual viscometric functions as $\sigma\equiv \Sigma'_{xy} = S\sin \theta$ and $N_1\equiv \Sigma'_{xx} - \Sigma'_{yy} = 2S\cos \theta$.

The major advantage of the polar representation is that it decouples the role of the two competing processes at stake, namely elastic relaxation and advection. 
Indeed, the amplitude of the microstructure anisotropy $S/\mu$ relaxes elastically, with a typical timescale $\tau_0$, while the tilt $\theta/2$ evolves only under advection, with a typical timescale $\dot\gamma^{-1}$.
In the elastically dominated flow regime $\dot\gamma \tau_0\ll 1$ we consider here, it results in a separation of timescales: $S$ is a fast variable, $\theta$ a slow one.

A final remark here is that if $\theta=2\alpha$, that is, $\bm{\Sigma}'$ aligned with $\Einf$, the most (respectively least) compressed contacts are along the compressional (resp. elongational) direction of flow.
As we will see, the stress evolution in \eqref{eq:Sigmaprim} is such that in practice, for $0<\alpha<\pi/2$, the vorticity rotates the microstructure and therefore misaligns it with respect to $\Einf$ (i.e.~$\theta \neq 2 \alpha$).
This simple mechanism is giving rise to most of the rheological phenomena discussed in this work.

\section{Dependence on flow geometry}

\subsection{Yield: von Mises criterion}

The question of the yield of soft jammed suspensions (and yield stress fluids in general~\cite{varchanisTransitionSolidLiquid2020}) in arbitrary geometries has recently seen a surge of activity~\cite{zhangYieldingFlowSoftJammed2018,thompsonYieldStressTensor2018,de_cagny_yield_2019,thompsonRheologicalMaterialFunctions2020}. 
This renewed interest stems from the development of new experimental techniques to measure the relevant stress components at yield in extensional flows~\cite{zhangYieldingFlowSoftJammed2018,varchanisTransitionSolidLiquid2020}. 
Indeed, historically the yield shear stress under simple shear has been by far the most commonly measured yield stress, and served as a reference point for the development of constitutive models. 

The motivation to compare the yield in extension and simple shear flows revolves around two questions: (i) the validity of the so-called von Mises criterion, which states that under an arbitrary deformation, yield occurs when 
$\sqrt{\bm{\Sigma}':\bm{\Sigma}'/2} = \tau_\mathrm{c}$, with $\tau_\mathrm{c}$ a geometry independent ``yield'' stress, 
and (ii) the weight of normal stress differences under simple shear in the von Mises criterion if it applies~\cite{thompsonYieldStressTensor2018}. 
Both issues remain unresolved, 
with experiments showing apparently contradicting results. 
Some experiments find a good agreement with the von Mises criterion~\cite{ovarlezThreedimensionalJammingFlows2010,shaukatShearMediatedElongational2012,de_cagny_yield_2019} while others do not~\cite{zhangYieldingFlowSoftJammed2018,thompsonRheologicalMaterialFunctions2020}. 
However in many works normal stress differences in simple shear are assumed to vanish, which is debated~\cite{thompsonYieldStressTensor2018,thompsonRheologicalMaterialFunctions2020}.

In our constitutive model, \eqref{eq:Sigmaprim}, the von Mises criterion naturally holds, 
with $\tau_\mathrm{c}(\phi) = \sqrt{\beta(\phi)/2\xi(\phi)}$. It shares this property with phenomenological models~\cite{oldroydRationalFormulationEquations1947,saramito_new_2007,saramito_new_2009,dimitriou_canonical_2019,saramitoNewBrittleelastoviscoplasticFluid2021}, 
while in MCT-ITT for soft glasses deviations from the von Mises criterion are tiny, despite the criterion not being baked in~\cite{brader_glass_2009}.

We have already shown that in simple shear, our model predicts that at yield the normal stress difference is larger in magnitude than the shear stress~\cite{CunyPRL21}. The contribution of normal stresses to the yield criterion thus cannot be neglected, 
which is consistent with some~\cite{zhangYieldingFlowSoftJammed2018,thompsonRheologicalMaterialFunctions2020}, but not all~\cite{de_cagny_yield_2019} experimental observations.
We here report the evolution of the relative contributions of shear and normal stresses to the von Mises criterion by exploring the entire family of possible flows for our constitutive model, given by \eqref{eq:flow_family}.

\begin{figure}
    \centering
  \includegraphics[width=\linewidth]{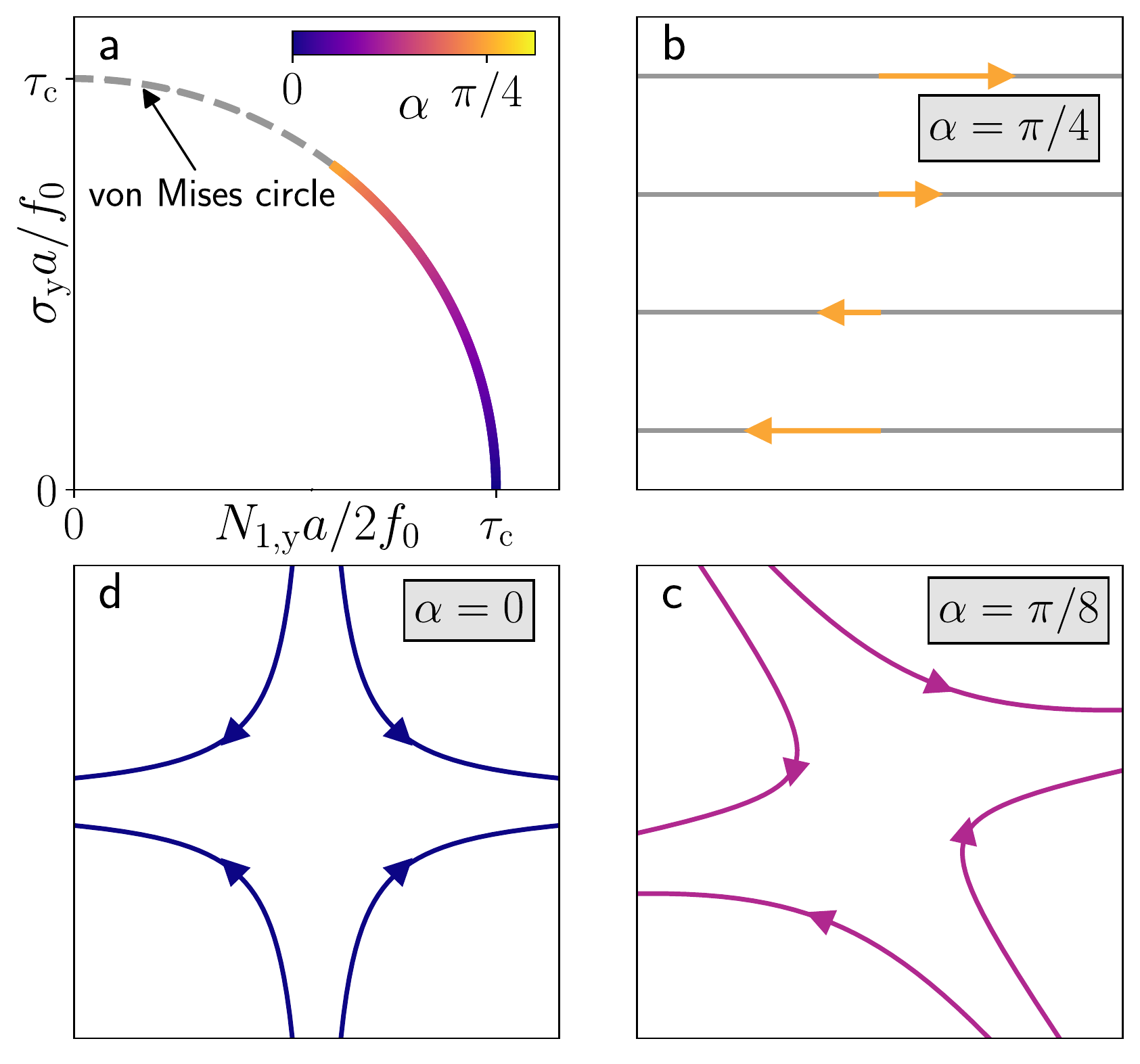}
  \caption{(a) The von Mises criterion (dashed gray line) in our model. For two-dimensional flows (parametrized by the angle $\alpha$ as defined in \eqref{eq:flow_family}), the dynamic yield stress is unique to each flow geometry, as shown in colored line. 
  (b)--(d) Different flows in the family, in clockwise order $\alpha=\pi/4$ (b, simple shear), $\alpha=\pi/8$ (c) and  $\alpha=0$ (d, planar extension).}
  \label{fig:vonMises}
\end{figure}

For a two-dimensional system, the von Mises criterion states that yield occurs on a circle of radius $\tau_\mathrm{c}$ in the plane $(N_1/2 ,\sigma)$, represented in \figref{fig:vonMises}{a} for $\phi=1.26$.
In our model, the degeneracy is lifted only for the dynamic yield stress tensor by taking the solution of \eqref{eq:Sigmaprim} in the $\dot\gamma\to 0$ limit. 
We thus here focus on the dynamic yield stress.
Defining the angle $\theta_\mathrm{y}$ separating the yield point from the $N_1/2$ axis, we have
\begin{equation}
    \theta_\mathrm{y} = 2\alpha - \arcsin\left[\frac{2\tau_\mathrm{c}(\phi)\tan\alpha}{\kappa(\phi)}\right]\, .
\end{equation}
This admits solutions only when $\alpha\in [0, \arctan\{\kappa(\phi)/2\tau_\mathrm{c}(\phi)\}]$. For larger values of $\alpha$, corresponding to flows mostly (but not purely) rotational, our constitutive equation admits no steady solution under flow, instead the stress tensor keeps rotating with the applied flow.
We represented the solution in \figref{fig:vonMises}{a}, along with some of the corresponding flows in panels (b)--(d).

Because $\theta_\mathrm{y}<\pi/2$ for all $\alpha$ values, normal stresses always have a significant contribution to the von Mises criterion.
This in particular points to the importance of measuring normal stresses at yield under simple shear for soft jammed suspensions~\cite{thompsonYieldStressTensor2018}.
Furthermore, in a three-dimensional system, a finite second normal stress difference would presumably also contribute to the von Mises criterion.

At the microstructure level, normal stresses in simple shear stem from the misalignment of the principal axes of the fabric tensor $\bm{Q}$ with the ones of the symmetrized strain rate tensor $\Einf$.
Due to the vorticity of the flow, the most compressed contacts lie between the compression and flow gradient directions. 
In consequence, the largest contact forces have a larger projection on the flow gradient direction than on the flow direction, which leads to $N_1>0$.

This challenges the common assumption that the deviatoric stress is  quasi-Newtonian~\cite{coussotGravityFlowInstability2005,castroComparisonMethodsMeasure2010,ovarlezThreedimensionalJammingFlows2010,saramitoProgressNumericalSimulation2017,zhangYieldingFlowSoftJammed2018}.
The main corollary of this assumption is that the sole (or at least the main~\cite{de_cagny_yield_2019}) contributor to the von Mises criterion 
is the projection of the deviatoric stress tensor on $\Einf$, i.e.~$\bm{\Sigma}':\Einf/\sqrt{2\Einf:\Einf}$~\cite{ovarlezThreedimensionalJammingFlows2010}.
Our constitutive model shows that this assumption is likely not verified in most flows except the most extensional ones. 
As soon as the vorticity tilts $\bm{Q}$ (and thus $\bm{\Sigma}'$) by an angle $\nu$ with respect to $\Einf$,   the component of $\bm{\Sigma}'$ tensorially orthogonal to $\Einf$ starts to contribute to $\bm{\Sigma}':\bm{\Sigma}'$.
For small $\nu$, this contribution is quadratic, but already for $\nu\approx\SI{9}{\degree}$ it accounts for $\SI{10}{\percent}$ of $\bm{\Sigma}':\bm{\Sigma}'$, and for $\nu\approx\SI{22}{\degree}$ it reaches $\SI{50}{\percent}$.

The role of the vorticity in a stress evolution such as \eqref{eq:Sigmaprim} is constrained by frame indifference for inertialess systems~\cite{nollContinuitySolidFluid1955}. 
It is thus generic that for constitutive models in the form of a stress evolution, the vorticity tends to give rise to a significant first normal stress difference in simple shear. 
This includes several popular phenomenological models for yield stress fluids~\cite{saramito_new_2007,saramito_new_2009,dansereauContinuumViscousElasticBrittleFinite2021}.
(A recent microstructure-based model for dense hard particle suspensions also falls into this class and similarly predicts significant normal stress differences~\cite{gillissenConstitutiveModelTimeDependent2019}.) 
The effect of the vorticity on the first normal stress difference can however still be mitigated in phenomenological models by using large prefactors in front of the stress relaxation term and/or the $\Einf$ source term.

Finally, it should be noted that our analysis focuses on the dynamic yield stress tensor. 
Although it is reasonable to believe that the static stress tensor will not be aligned with $\Einf$ if the dynamic one is not, it is also possible that the effect of the vorticity is less pronounced on the former, as static yield occurs after a finite strain of order \numrange{0.1}{1}, whereas the dynamic yield stress is measured in the limit of infinite strain.

\subsection{Steady flow: extensional vs simple shear}

We now turn to steady flow, focusing on simple shear flow ($\alpha=\pi/4$)  with rate $\dot\gamma$
\begin{equation}
    \nabla \bm{u}^\infty = \begin{pmatrix} 
                            0 & \dot\gamma\\
                            0 & 0
                            \end{pmatrix}\, ,
\end{equation}
and planar extensional flow ($\alpha=0$)  with rate $\dot\epsilon$ (note that $\dot\epsilon = \dot\gamma/2$)
\begin{equation}
    \nabla \bm{u}^\infty = \begin{pmatrix} 
                            \dot\epsilon & 0\\
                            0            & -\dot\epsilon
                            \end{pmatrix}\, .
\end{equation}

Specifying \eqref{eq:Sigmaprim} for a simple shear, we get 
\begin{equation}
  \label{eq:shear_cart}
  \begin{aligned}
    \dot\sigma &  = \frac{\kappa-N_1}{2}\gd + \left[\beta-2\xi\left(N_1^2/4+\sigma^2\right)\right]\sigma\, , \\
    \dot N_1 & =  2\gd \sigma  + \left[\beta-2\xi\left(N_1^2/4+\sigma^2\right)\right]N_1\, .
  \end{aligned}
\end{equation}
Similarly, in planar extension, we get
\begin{equation}
  \label{eq:extensional_cart}
  \begin{aligned}
  \dot\sigma &  = \left[\beta-2\xi\left(N_1^2/4+\sigma^2\right)\right]\sigma\, , \\
    \dot N_1 & =  2 \dot\epsilon \kappa  + \left[\beta-2\xi\left(N_1^2/4+\sigma^2\right)\right]N_1.
\end{aligned}
\end{equation}

\begin{figure}
  \centering
   \includegraphics[width=0.9\columnwidth]{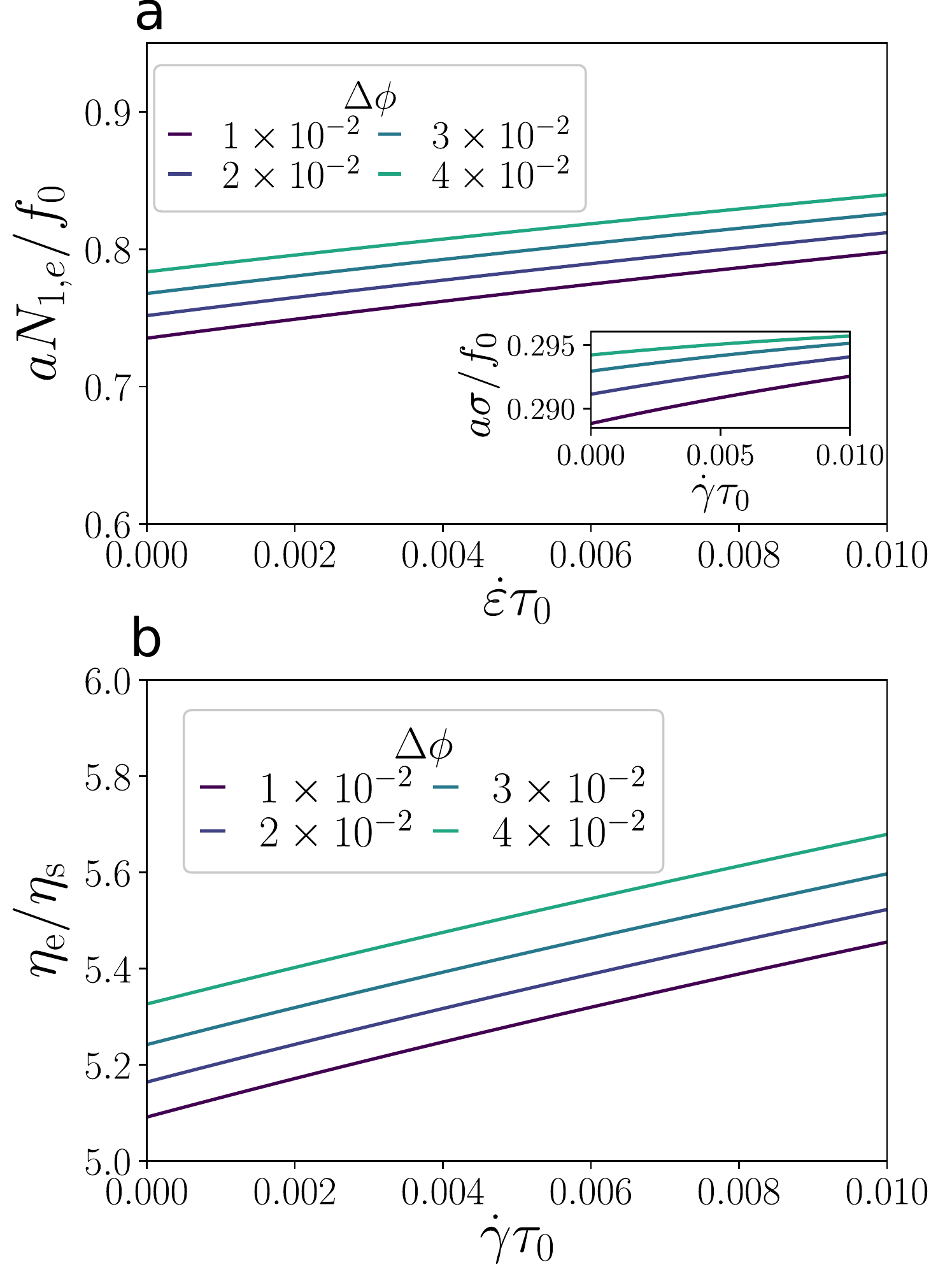}
  \caption{(a) Flow curves in planar extensional flow ($\alpha = 0$): normal stress difference as a function of the extensional shear rate, for several volume fractions above jamming, labelled by their distance to jamming $\Delta \phi = \phi - \phi_\mathrm{J}$. In inset, flow curves in simple shear ($\alpha=\pi/4$), for the same volume fractions. (b) Corresponding Trouton ratio as a function of the applied shear rate.}
  \label{fig:flow_curves}
\end{figure}

The steady state flow curves are given by the stable fixed point of these equations.
The simple shear case has been exposed in~\cite{CunyPRL21}, but we replot it in the inset of \figref{fig:flow_curves}{a}, alongside the predicted rheology for extensional flow, to ease the comparison. 
As already discussed in a previous section, under both deformations the model shows a yield stress, for simple shear on both $\sigma$ and $N_1$, and in extensional flow only on $N_1$. 
Unsurprisingly, by symmetry $\sigma$ vanishes in steady extensional flow. This can also be seen in  \eqref{eq:extensional_cart}, as we always have $\left[\beta-2\xi\left(N_1^2/4+\sigma^2\right)\right]<0$ above the normal yield stress.
None of the flow curves is following a Herschel-Bulkley law, but rather a Bingham law at small shear rates.
Indeed, expanding the stable stationary solutions of \eqref{eq:shear_cart} for $N_1$ and $\sigma$ up to first order in $\gd$, we get the following expression for stationary state under simple shear
\begin{align}
   & \sigma = \text{sgn}(\gd)\, \sigma_\mathrm{y,s} + \left(\frac{\kappa}{4\beta}-\frac{1}{\kappa\xi}\right)\gd + o(\gd), \\
   & N_1 = N_\mathrm{y,s} + \frac{2\sy}{\beta} \, |\gd| + o(\gd),
\end{align}
with $N_\mathrm{y,s}=2\beta/(\xi\kappa)$ and $\sigma_\mathrm{y,s}=\sqrt{\beta\left[1-2\beta/(\xi\kappa^2)\right]/(2\xi)}$.
Similarly, from \eqref{eq:extensional_cart}, we get the flow curve at first order under planar extension
\begin{align}
  \label{exp:extensional_ss_N}
  N_1 = 2\text{sgn}(\dot\epsilon) \tau_\mathrm{c} + \frac{\kappa}{\beta}\dot\epsilon + o(\dot\epsilon).
\end{align}

From these curves, we can extract the Trouton ratio predicted by our model. 
The Trouton ratio for planar extension is defined as $\mathrm{Tr} = \eta_\mathrm{e}/\eta_\mathrm{s}$, with $\eta_\mathrm{e} = (\Sigma'_{11}-\Sigma'_{22})/\dot\epsilon$ in planar extension and $\eta_\mathrm{s} = \Sigma'_{12}/\dot\gamma$ in simple shear, evaluated at $\dot\gamma=\dot\epsilon$. 
For a Newtonian fluid, the Trouton ratio takes a value of $4$.
We show in \figref{fig:flow_curves}{b} that our model predicts a Trouton ratio quite larger than the Newtonian value for soft jammed suspensions.
This is another manifestation of the significant rotation of the microstructure and stress tensors under the action of the vorticity in our model.
This result stands in contrast with MCT for soft glasses, which predicts a sub-Newtonian Trouton ratio~\cite{braderFirstPrinciplesConstitutiveEquation2008}. 
Experimentally, Trouton ratios significantly larger than the Newtonian value have been observed for uniaxial elongation of concentrated emulsions~\cite{anklamUseOpposedNozzles1994,rozanskaExtensionalViscosityMeasurements2014}.

\section{Transients}

In this section we consider the time evolution of the stress provided in \eqref{eq:Sigmaprim} under time-dependent driving protocols 
which are commonly explored experimentally.

\begin{figure}
  \centering
    \includegraphics[width=0.9\columnwidth]{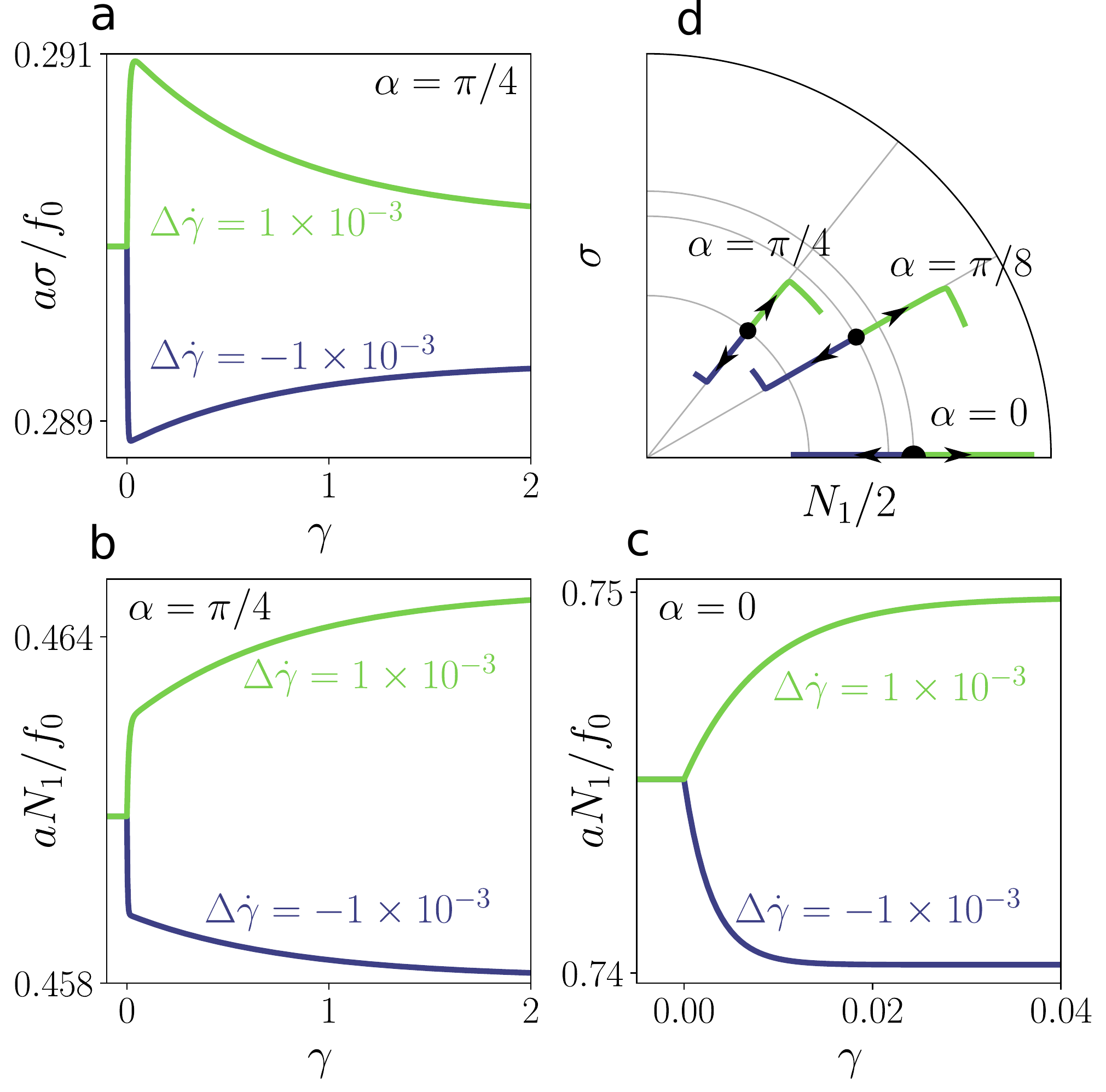}
  \caption{
(a) Shear stress as a function of the applied strain after respectively step increase (green curve) and step decrease (blue curve) of the shear rate, with $\dot\gamma_0 \tau_0 = \num{2e-3}$, $\Delta\dot\gamma\tau_0 = \pm \num{e-3}$ and $\Delta\phi=\num{e-2}$, under simple shear. (b) Associated normal stress difference versus strain, under the same conditions as (a). (c) Normal stress difference as a function of the applied strain in planar extensional flow, for otherwise same conditions as in (a)-(b). 
(d) Schematic trajectories in the $(N_1/2, \sigma)$ plane in simple shear ($\alpha =\pi/4$), planar extension ($\alpha = 0$), and a flow in between ($\alpha=\pi/8$). Overshoots and undershoots for $\alpha \neq 0$ come from the radial dynamics being much faster than the azimuthal one.
}
  \label{fig:step_rate}
\end{figure}

\subsection{Steps in rate: stress overshoot, stress relaxation}

We will here focus on the step-in-shear-rate protocol. 
Starting from a steady state under a shear rate $\dot\gamma_0$, one suddenly increases or decreases the  applied shear rate to $\dot\gamma_0 + \Delta \dot\gamma$.
For this protocol, it is convenient to consider the dynamics in polar coordinates $S$ and $\theta$ (see \eqref{eq:polar_Sigma}). From \eqref{eq:Sigmaprim}, we have
\begin{align}
    \dot{S} &= \frac{\dot\gamma \kappa}{2}\cos (2\alpha -\theta) + (\beta - 2\xi S^2)S \\
    \dot{\theta} & = \dot\gamma \left[\frac{\kappa}{
    2S} \sin (2\alpha - \theta) - \tan\alpha\right]\, .
    \label{eq:constitutive_polar}
\end{align}
As discussed in introduction, in the elastic flow regime $S$ has a much faster dynamics than $\theta$.
This is clearly illustrated here by the decompositions $S = S_0 + \delta S$ and $\theta = \theta_0 + \delta \theta$, with $S_0$ and $\theta_0$ the steady-state values at $\dot\gamma_0$. 
From \eqref{eq:constitutive_polar} we see that right at the step in shear rate $\delta\dot{S} \sim \Delta\dot\gamma \tau_0$, 
whereas $\delta \dot\theta = 0$, while just after the increment, $\delta \dot\theta \sim \dot\gamma \tau_0 \delta S/S_0$.
We represented this dynamics in the polar plane $(S, \theta)$ in \figref{fig:step_rate}{d}, 
illustrating the fast radial relaxation and the comparatively slow azimuthal one, irrespective of the value of $\alpha$ (with the exception of $\alpha=0$, for which the $\theta$ dynamics is frozen).
From this representation we can easily predict the transients of $\sigma = S \sin \theta$ and $N_1=2S\cos\theta$, which we represent in \figref{fig:step_rate}{a}.

An increase in shear rate ($\Delta \dot\gamma>0$) will trigger a fast increase of $S$ and a slow decrease of $\theta$, leading to an overshoot in $\sigma$ and a monotonic increase in $N_1$.
Conversely, a decrease in shear rate ($\Delta \dot\gamma<0$) will lead to an undershoot in $\sigma$ and a monotonic decrease in $N_1$.
Integrating the dynamics numerically, we indeed confirm this behavior in~\figref{fig:step_rate}{a-b} for $\alpha=\pi/4$ and $\Delta\dot\gamma = \pm \num{e-3} $.
Of course the planar extension ($\alpha=0$) here stands out, as $\sigma=0$ and the relaxation proceeds only on $N_1$ and is monotonic, as shown in~\figref{fig:step_rate}{c}.

A particular case of step-change-in-shear-rate is flow cessation, i.e.~$\Delta \dot\gamma = -\dot\gamma_0$. 
On flow cessation in simple shear the shear stress relaxes down to a finite value, the residual stress $\sigma_\mathrm{r}$. 
An intriguing observation is that $\sigma_\mathrm{r}$ is a decreasing function of $\dot\gamma_0$; the faster one initially shears, the smaller the residual stress on cessation~\cite{mohan_microscopic_2013,mohan_build-up_2014,lidon_power-law_2017,vasishtResidualStressAthermal2021}.
While this effect may involve a stress-dependent cooperativity in the plasticity (e.g. plastic avalanches), we have shown that it can already be qualitatively reproduced by our model~\cite{CunyPRL21}.
The mechanism can also be easily understood in the polar representation. According to \eqref{eq:constitutive_polar}, when the shear is stopped $\theta$ is frozen and only $S$ relaxes down to the von Mises yield value $S = \sqrt{\beta/2\xi}$. 
Because the steady-state value of $\theta$ is a decreasing function of the shear rate for any $\alpha>0$, so is the residual stress $\sigma_\mathrm{r} =\sqrt{\beta/2\xi}\sin \theta $. 
This is true whenever $0<\alpha<\pi/2$, that is, as long as there is a finite vorticity in presence of finite straining in the flow geometry.

\subsection{Imposed stress: creep}

\begin{figure}
  \centering
  \includegraphics[width=0.99\linewidth]{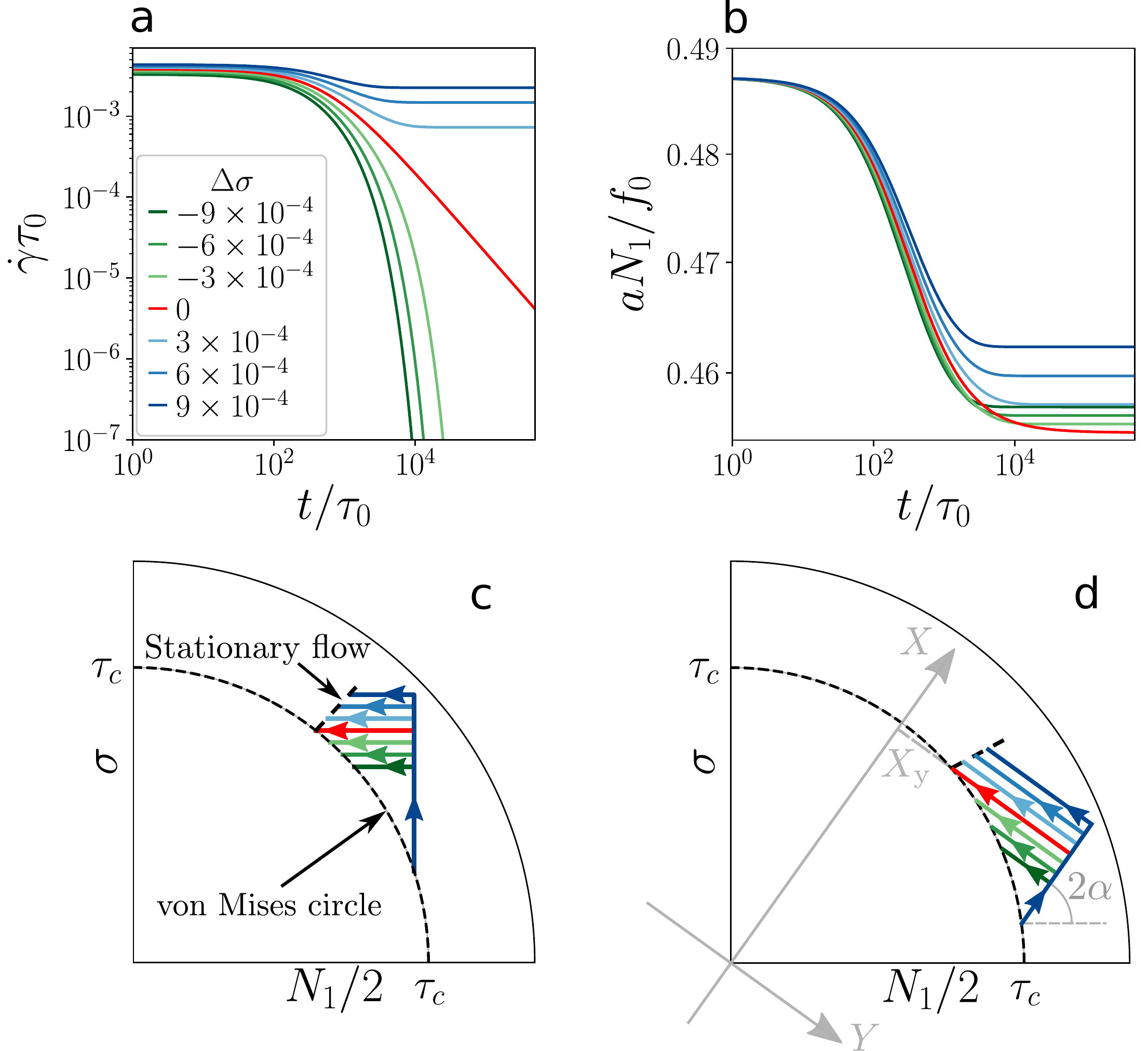}
  \caption{(a-b) Creep flow in simple shear for several values of imposed shear stresses $\sigma_0=\sy+\Delta\sigma$, with $\Delta\phi=0.01$, after a pre-shear under shear rate $\gps\tau_0=10^{-2}$. Shear rate (a) and normal stress difference (b) as a function of time. (c) Schematic trajectories in the $(N_1/2, \sigma)$ plane for creep in simple shear, showing the origin of the minimum in the long-time normal stress difference as a function of $\Delta \sigma$. (d) Schematic trajectories in the $(N_1/2, \sigma)$ plane for creep under flow with generic $\alpha$ value. 
  Creep corresponds to flow under imposed $X=\bm{\Sigma}':\Einf/\dot\gamma$.}
  \label{fig:creep}
\end{figure}

We now turn to the behavior under constant applied stress.
We follow here the protocol of Ref.~\cite{lidon_power-law_2017}.
We first apply a pre-shear under a large shear rate, 
and then let the stress relax to its residual stress tensor $\bm{\Sigma_\mathrm{r}}$ under vanishing shear. 
From this initial state, we perform a shear under constant stress.

The stress component we impose depends on the flow. 
For simple shear flow, we impose the shear stress, which from \eqref{eq:shear_cart} with $\dot{\sigma} = 0$ results in a shear rate
\begin{align}
  \label{eq:gd_N1_creep}
  \gd=\frac{2\sigma_0}{N_1-\kappa}\left[\beta - 2\xi\left(N_1^2/4 + \sigma_0^2\right)\right]\, .
\end{align}
Replacing $\gd$ by its expression in the evolution equation for $N_1$, we obtain the following dynamics for $N_1$:
\begin{equation}
  \label{eq:N1_creep}
  \dot{N}_1 = \left(\frac{4\sigma_0^2}{N_1-\kappa}+N_1\right)\left[\beta - 2\xi\left(N_1^2/4 + \sigma_0^2\right)\right].
\end{equation}
In~\figref{fig:creep}{a}, we show numerical integrations of these equations for several values of the imposed shear stress $\sigma$, ranging from below to above the simple shear yield stress $\sigma_\mathrm{y,s}$.
Predictably, for $\sigma<\sigma_\mathrm{y,s}$, the flow halts after a finite time, while for $\sigma>\sigma_\mathrm{y,s}$ the shear rate decreases  towards its finite steady-state value.
The normal stress difference $N_1$, however, always decays to a finite value, see~\figref{fig:creep}{b}. 
This value is moreover a non-monotonic function of $\sigma$ taking a minimum for $\sigma=\sigma_\mathrm{y,s}$. 
This is most easily understood from the loci of fixed points in the $(N_1/2, \sigma)$ plane, represented in~\figref{fig:creep}{c}.
For $\sigma<\sigma_\mathrm{y,s}$, fixed points all lie on the von Mises yield circle ($S=\tau_\mathrm{c}$), with an azimuth $\theta$ increasing (and thus $N_1=2S \cos\theta$ decreasing) with increasing $\sigma$.
For $\sigma>\sigma_\mathrm{y,s}$, $S$ increases with $\sigma$ fast enough to counteract the continuing increase of $\theta$, so that $N_1$ now increases with $\sigma$.

For $\sigma=\sigma_\mathrm{y,s}$, we get power law decay of the shear rate. 
A quick look at \eqref{eq:N1_creep} reveals that for $\sigma=\sigma_\mathrm{y,s}$ the fixed point of the $N_1$ dynamics (located at $N_{1,\mathrm{y,s}}$) is marginally stable, so that the late dynamics of $\delta N_1 = N_1-N_{1,\mathrm{y,s}}$ is $\delta\dot{N}_1 \propto -(\delta N_1)^2$. 
This implies that at late times $\delta N_1\propto t^{-1}$, which in turn, from \eqref{eq:gd_N1_creep} implies that the final decay of the shear rate is in $\dot\gamma \propto t^{-1}$, which is indeed the exponent we observe in \figref{fig:creep}. 
A power law decay $\dot\gamma \propto t^{-b}$ is also observed experimentally~\cite{divouxStressinducedFluidizationProcesses2011,lidon_power-law_2017}, but with $b \approx 2/3$.
A recent phenomenological constitutive model showing otherwise good agreement with experimental data on microgels predicts $b=1/n>1$, with $n<1$ the Herschel-Bulkley exponent~\cite{kamaniUnificationRheologicalPhysics2021}.

Creep can also be defined for more general flows parametrized by $\alpha$ (see \eqref{eq:flow_family}), as a deformation under constant $X \equiv \bm{\Sigma}':\Einf/\dot\gamma$.
The quantity $X$ is the projection of the stress tensor on $\Einf$, and measures the dissipative part of the stress~\cite{giusteriTheoreticalFrameworkSteadystate2018}. 
It reduces to $\sigma$ in simple shear.
In the plane $(N_1/2, \sigma)$, stresses sharing the same $X$ value fall on a line making an angle $2\alpha - \pi/2$ with the horizontal axis, as depicted in Fig.~\ref{fig:creep}d.
Creep for arbitrary $\alpha$ follows the same phenomenology than in simple shear.
For $X<X_\mathrm{y}$, with $X_\mathrm{y}$ the yield dissipative stress, the system quickly relaxes towards an arrested state. 
For $X = X_\mathrm{y}$, the shear rate decays at long times as $t^{-1}$.
For  $X>X_\mathrm{y}$, the system reaches a steady flow.
Separately, the non-dissipative part of the stress $Y$ (which we can define as $Y \equiv \sqrt{|\bm{\Sigma}' - 2X\Einf|^2/2}$) relaxes towards a value which is a non-monotonic function of the imposed $X$, taking a minimum for $X=X_\mathrm{y}$.

\section{Parallel superposition}

\begin{figure}
  \centering
  \includegraphics[width=0.99\columnwidth]{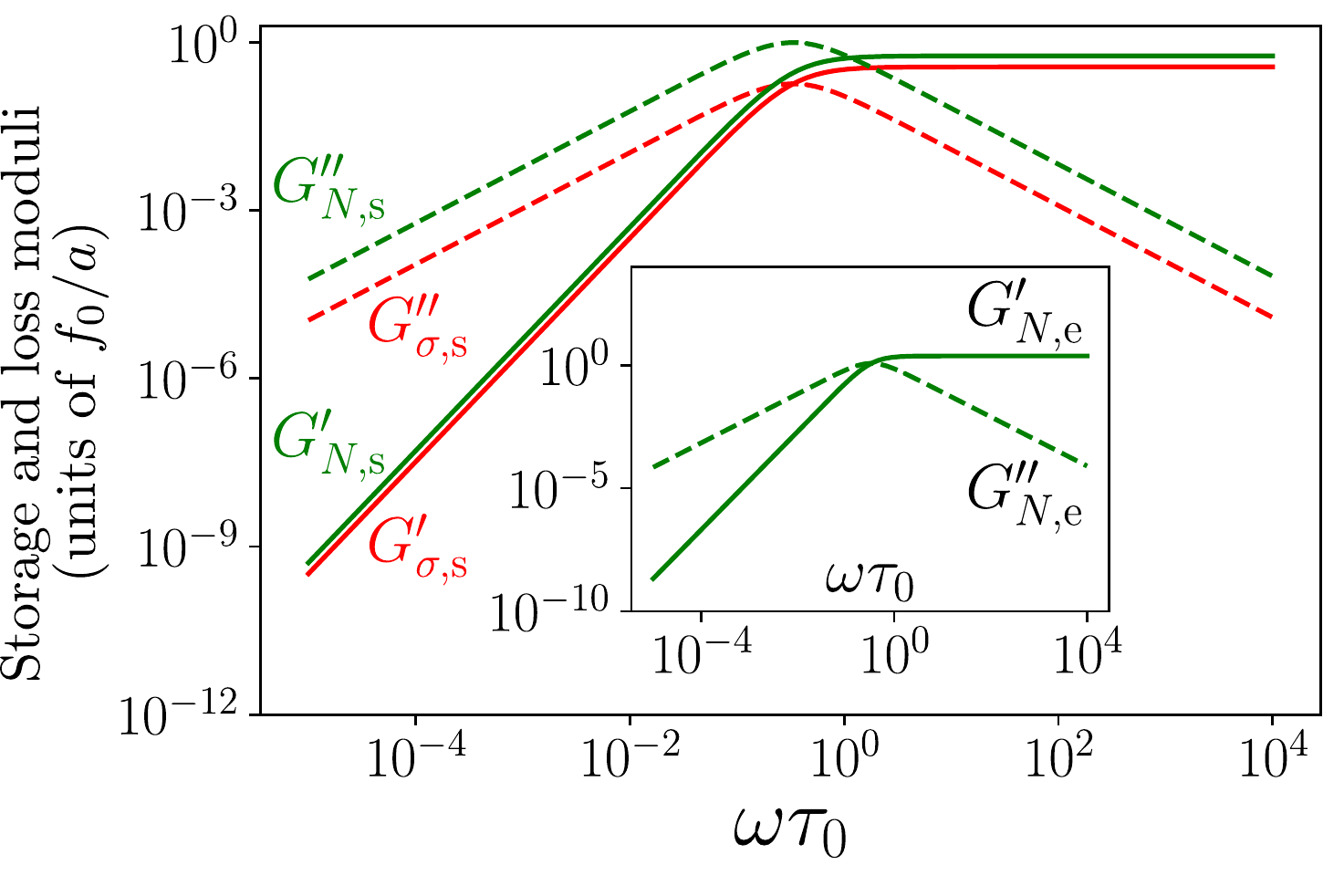}
  \caption{Frequency dependence of the storage and loss moduli under parallel superposition on the von Mises yield circle in simple shear (main panel) and in planar extension (inset).}
  \label{fig:modules}
\end{figure}

The question of the visco-elasticity of flowing soft jammed materials has  received some attention recently~\cite{benmouffok-benbelkacemNonlinearViscoelasticityTemporal2010,ngouambaElastoplasticBehaviorYield2019}.
In particular, a central question is the quantification of  elasticity in yielded suspensions, which are not expected to be purely viscous.
We here report the predictions of our constitutive model, \eqref{eq:Sigmaprim}, under parallel superposition. 
We follow a protocol similar to the one of Ref.~\cite{ngouambaElastoplasticBehaviorYield2019}, and apply in simple shear a shear stress with a small oscillation around an averaged value, that is, $\sigma=\bar{\sigma}+\mathrm{Re}(\delta\sigma\, e^{i\omega t})$ with $\bar{\sigma}>\sigma_\mathrm{y,s}$. We then measure a response in shear rate $\dot\gamma = \bar{\dot\gamma} + \mathrm{Re}(i \omega \delta{\gamma} e^{i\omega t})$, and normal stress difference $N_1=\bar{N}_1+\mathrm{Re}(\delta N_1 \, e^{i\omega t})$. The DC components then satisfy
\begin{align}
    &\bar{N}_1=\frac{\kappa-\sqrt{\kappa^2-16\bar{\sigma}^2}}{2} \\
    &\bar{\gd}=\frac{2\bar{\sigma}}{\bar{N}_1-\kappa}\left[\beta - 2\xi\left(\frac{\bar{N}_1^2}{4} + \bar{\sigma}^2\right)\right]\,.
\end{align}

Linearizing \eqref{eq:shear_cart}, one finds that $\delta{\gamma}$, $\delta{\sigma}$ and $\delta N_1$ satisfy
\begin{equation}
\begin{split}
  \label{eq:shear_cart_osc_tot}
    &\left[i\omega + \frac{E_\sigma}{\eta_\sigma}\right]\delta\sigma   = i\omega E_\sigma \delta\gamma + C_\sigma \delta N_1\, , \\
    & \left[i\omega+\frac{E_N}{\eta_N}\right]\delta N_1 = 
    i\omega E_N \delta\gamma + C_N \delta\sigma\, ,
\end{split}
\end{equation}
with effective elastic moduli and viscosities $E_\sigma = (\kappa - \bar{N}_1)/2$,
$\eta_\sigma = (\kappa - \bar{N}_1)/[-2\beta +\xi (\bar{N}_1^2 + 12\bar{\sigma}^2)]$,
$E_N = 2\bar{\sigma}$ and $\eta_N = 2\bar{\sigma}/[-\beta + \xi (3\bar{N_1}^2/2 + 2\bar{\sigma}^2)]$, and couplings $C_\sigma = - \left(\bar{\dot{\gamma}}/2 +\xi \bar{N}_1\bar{\sigma}\right)$ and $C_N = \left(2\bar{\dot{\gamma}}-4\xi\bar{\sigma} \bar{N}_1\right)$.

Thus at linear level (\eqref{eq:shear_cart_osc_tot}), under parallel superposition our model can be interpreted as a pair of Maxwell models for shear and normal stresses, coupled through $C_\sigma$ and $C_N$, only with effective viscosities and elastic moduli which depend on the mean applied stress $\bar{\sigma}$. 
From \eqref{eq:shear_cart_osc_tot}, we can get the storage and loss moduli 
$G'_{\sigma, \rm s} = \mathrm{Re}(\delta\sigma/\delta \gamma)$ and 
$G''_{\sigma, \rm s} = \mathrm{Im}(\delta\sigma/\delta \gamma)$, 
as well as equivalent moduli for normal stresses 
$G'_{N, \rm s} = \mathrm{Re}(\delta N_1/\delta \gamma)$ and 
$G''_{N, \rm s} = \mathrm{Im}(\delta N_1/\delta \gamma)$.

\begin{figure*}[h]
  \centering
\includegraphics[width=\textwidth]{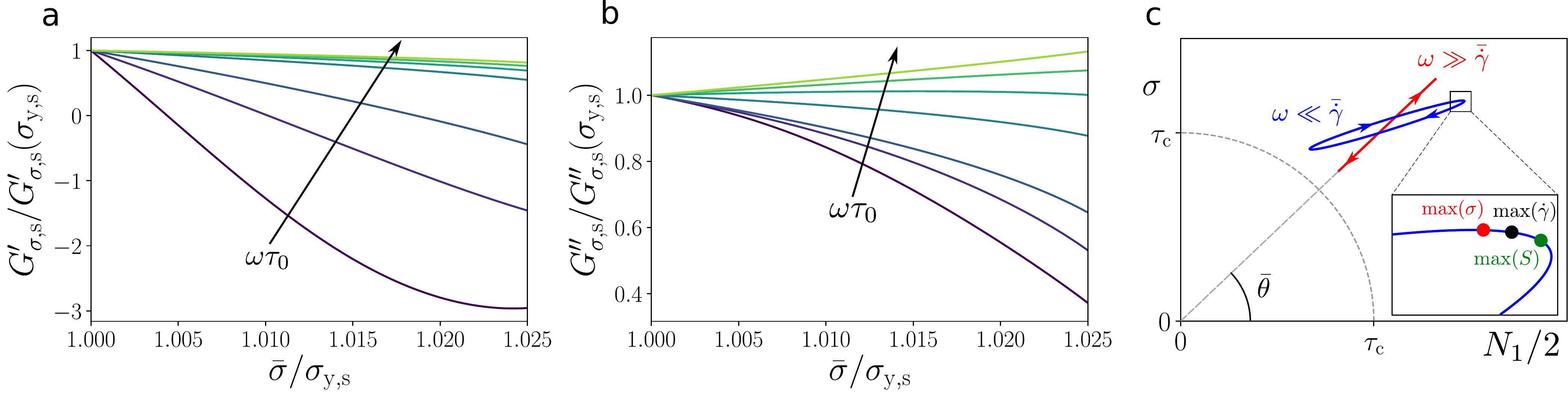}
  \caption{Storage (a) and loss (b) moduli predicted for a parallel superposition as a function of $\bar{\sigma}\geq \sigma_{\mathrm{y,s}}$ in simple shear for $\Delta\phi=\num{e-2}$. Curves from bottom to top correspond to  $\omega\tau_0=\num{3e-2}$, $\num{5e-2}$, $\num{8e-2}$, $\num{3e-1}$, $\num{5e-1}$, $\num{7e-1}$, and $1$. (c) Schematic trajectories in the $(N_1/2, \sigma)$ plane at high (red segment) and low (blue ellipse) frequencies. In inset, a zoomed in view on the apex of the ellipse followed clockwise at low frequencies, which shows that the maximum of $\sigma$ occurs before the maximum of $\dot\gamma$ (which in turn precedes the maximum of $S$), giving rise to a negative storage modulus $G'_{\sigma, \mathrm{s}}$.}
  \label{fig:linear_superposition}
\end{figure*}

These are plotted as a function of frequency at yield, i.~e. for $\bar{\sigma} = \sigma_\mathrm{y,s}$ in  \figref{fig:modules}.
The response is unsurprisingly the one of a Maxwell model, with a material essentially viscous at low frequencies and elastic at high frequencies. 
Indeed, assuming that $\bar{\sigma} = \sigma_\mathrm{y,s}$, a short calculation leads to
\begin{align}
G'_{\sigma, \rm s} & = \frac{\left(\kappa-N_\mathrm{y,s}\right)\omega^2}{2(\omega^2+4\beta^2)},
  && G''_{\sigma, \rm s} = \frac{\left(\kappa-N_\mathrm{y,s}\right)\beta\omega}{\omega^2+4\beta^2}\, , \\
  G'_{N, \rm s} & = \frac{2\sigma_\mathrm{y,s}\omega^2}{\omega^2+4\beta^2}, &&  G''_{N, \rm s}= \frac{4\beta\omega}{\omega^2+4\beta^2}\, .
\end{align}
The frequency dependence is not reported in the experimental literature, either because it is not accessible with the employed technique~\cite{benmouffok-benbelkacemNonlinearViscoelasticityTemporal2010}, or because it is argued to be mild~\cite{ngouambaElastoplasticBehaviorYield2019}.

The same analysis can be carried out in any flow geometry, and for instance we show in the inset of Fig.~\ref{fig:modules}
the visco-elasticity at yield under planar extension, which involves only the storage and loss moduli associated with the normal stress difference
\begin{equation}
  G'_{N, \rm e}  = \frac{2\kappa\omega^2}{\omega^2+4\beta^2}, \qquad  \qquad G''_{N, \rm e}  = \frac{4\kappa\beta\omega}{\omega^2+4\beta^2}.
\end{equation}

Coming back to visco-elasticity in simple shear, we show the storage and loss moduli as a function of mean stress $\bar{\sigma}$ for several frequencies in~\figref{fig:linear_superposition}.
The storage modulus is in qualitative agreement with experiments, with $G'_{\sigma, \rm s}$ rapidly decreasing just above yield for low frequencies~\cite{benmouffok-benbelkacemNonlinearViscoelasticityTemporal2010,ngouambaElastoplasticBehaviorYield2019}. 
At low frequencies, the loss modulus decreases with increasing $\bar{\sigma}$, but at higher frequencies (for the volume fraction shown in \figref{fig:linear_superposition}, for $\omega\tau_0\gtrsim \num{5e-1}$), the trend changes and $G''_{\sigma, \rm s}$ becomes an increasing function of $\bar{\sigma}$.
Experimentally, Ref.~\cite{ngouambaElastoplasticBehaviorYield2019} reports a decrease of $G''_{\sigma, \rm s}$ with $\bar{\sigma}$.
From the data in Ref.~\cite{benmouffok-benbelkacemNonlinearViscoelasticityTemporal2010}, assuming that the coefficients of the Jeffrey model used in this work are frequency independent, we can also infer a decrease of $G''_{\sigma, \rm s}$ with $\bar{\sigma}$, however this decrease gets milder with increasing frequencies. 
This suggests that the role of frequency is perhaps overlooked in the literature.

An apparently quite surprising result is that the storage modulus $G'_{\sigma, \rm s}$ can become negative when $\bar{\sigma}$ increases. 
Said otherwise, the suspension seemingly behaves like a visco-elastic fluid with \emph{negative} elastic modulus.
To understand this result, it is useful to return to the polar representation of the stress in \eqref{eq:constitutive_polar}, where the fact that the dynamics of $\theta$ is much slower than the one of $S$ again plays a key role.
Introducing $S = \bar{S} + \delta S e^{i\omega t}$ and $\theta = \bar{\theta}+\delta \theta e^{i\omega t}$, $\delta S$ and $\delta \theta$ follow
\begin{equation}
    \begin{split}
    \left[i\omega - \beta + 6\xi \bar{S}^2\right] \delta S & = i\omega \frac{\kappa \sin \bar{\theta}}{2}\delta \gamma + \frac{\bar{\dot\gamma} \kappa \cos\bar{\theta}}{2}\delta \theta \\
    \left[i\omega +\dot\gamma \frac{\kappa}{2\bar{S}}\sin\bar{\theta}\right] \delta \theta & = -\frac{\bar{\dot\gamma} \kappa\cos\bar{\theta}}{2\bar{S}^2} \delta S
\end{split}\label{eq:linear_response_polar}
\end{equation}
We sketch in Fig.~\ref{fig:linear_superposition}c the dynamics in the $(S,\theta)$ plane under the two opposite regimes $\omega/\dot\gamma \ll 1$ and $\omega/\dot\gamma \gg 1$.
Under oscillatory shear at large frequencies $\omega/\bar{\dot\gamma} \gg 1$, we can neglect the terms involving $\bar{\dot\gamma}$ in \eqref{eq:linear_response_polar}.
We thus notice that $\theta$ is essentially frozen, and $S$ follows a usual Maxwell model 
\begin{equation}
    \left(i\omega + \frac{E_\mathrm{S}}{\eta_\mathrm{S}}\right) \delta S  = i\omega E_\mathrm{S} \delta \gamma\, ,
\end{equation}
with effective elastic modulus and viscosity
\begin{equation}
    E_\mathrm{S} =  \frac{\kappa \sin \bar{\theta}}{2}>0\, , \qquad \eta_\mathrm{S} =  \frac{\kappa \sin\bar{\theta}}{-2\beta +12\xi \bar{S}^2}>0\, .
\end{equation}
To determine the sign of $\eta_\mathrm{S}$, we used that $-2\beta +12\xi \bar{S}^2 > - 2\beta +12\xi \tau_\mathrm{c}^2 = 4\beta >0$.
In this limit the system thus follows a radial oscillation in the $(S, \theta)$ plane, such that $\delta \sigma$ is in phase with $\delta S$, leading to a positive $G'_\mathrm{\sigma, s}$.

In the $\omega \ll \dot\gamma$ regime, at linear order in $\omega$ the $S$ response in \eqref{eq:linear_response_polar} is still an effective Maxwell model, now 
with effective elastic modulus and viscosity
\begin{equation}
    E_\mathrm{S} =  \frac{\kappa \sin \bar{\theta}}{2\left[1-(\tan\bar{\theta})^{-2}\right]}\, , \qquad \eta_\mathrm{S} =  \frac{\kappa \sin\bar{\theta}}{-2\beta +12\xi \bar{S}^2 + \frac{\bar{\dot\gamma}\kappa}{\bar{S}} \cos\bar{\theta}} > 0\, .
\end{equation}
Now, $E_\mathrm{S}>0$ if $\bar{\theta}>\pi/4$, which is fulfilled within our model for small shear rates as $\cos \theta_\mathrm{y} = \kappa \sqrt{2\beta/\xi} > 1/\sqrt{2}$ for volume fractions close to jamming. 
$S$ therefore still responds like a usual visco-elastic variable, with positive storage modulus.
However, now $\theta$ also oscillates, 
so that in the $(S, \theta)$ plane, the system follows an elongated ellipse.
According to the second line of \eqref{eq:linear_response_polar}, $\delta\theta$ is almost in phase opposition with $\delta S$, but not quite: $S$ reaches its maximum slightly before $\theta$ hits its minimum point, the phase difference differing from $\pi$ by an angle of order $\omega/\bar{\dot\gamma}$.
This implies that the ellipse is followed in a clockwise manner.
This in turn implies that $\sigma$ reaches its maximum slightly before $S$ does, as can be easily understood from the inset of Fig.~\ref{fig:linear_superposition}c. 
More precisely, using $\delta \sigma = \sin \bar{\theta} \delta S + \cos\bar{\theta} \bar{S} \delta \theta$, we have
\begin{equation}
    \delta \sigma = \left(\frac{-\cos 2\bar{\theta}}{\cos \bar{\theta}} + \frac{2i\omega}{\bar{\dot\gamma} \kappa \tan\bar{\theta}}\right) \frac{\delta S}{\tan\bar{\theta}}
\end{equation}
which shows that the phase difference between $\delta S$ and $\delta\sigma$ is also of order $\omega/\bar{\dot\gamma}$.
It turns out that the values of $\bar{\theta}$ and $\kappa$ are such that the phase advance of $\delta \sigma$ on $\delta S$ is actually large enough to make the phase difference between $\delta \sigma$ and $\delta \gamma$ larger than $\pi/2$, which corresponds to a negative storage modulus $G'_{\sigma, \mathrm{s}}$.

Physically, the situation is thus clear. 
Strain rate oscillations induce both an oscillation in the overlap between particles in contact, but also in the orientation of these contacts because of the vorticity.
The negative storage modulus observed under flow is just a consequence of the oscillation of the principal axes of the microstructure  under slow enough  oscillatory driving, and more specifically of the small lag of the orientation of the contact with respect to the oscillation of the overlaps.

Our model thus offers a simple physical picture for the collapse of the storage modulus just above yield~\cite{benmouffok-benbelkacemNonlinearViscoelasticityTemporal2010,ngouambaElastoplasticBehaviorYield2019}, 
which differs significantly from the naive interpretation of a loss of elastic network, which is difficult to believe in a concentrated soft suspension above the jamming volume fraction.
It is actually much more plausible that the storage modulus becomes negative because of the rotation of the microstructure in near-antiphase with the driving oscillation, 
in a system with otherwise essentially intact elastic integrity (and this despite the contacts making up the elastic backbone being constantly renewed by the flow).
Once again, measuring the frequency dependence of the visco-elastic moduli seems essential in order to test these ideas in experiments.
 
Finally, let us remark that a very similar phenomenology is actually observed in polymer solutions~\cite{booijInfluenceSuperimposedSteady1966,kataokaInfluenceSuperimposedSteady1969,macdonaldParallelSuperpositionSimple1973}, with a negative storage modulus at low frequencies, and a simple tensorial Maxwell model including the vorticity in a frame-indifferent manner is known to recover qualitatively the behavior~\cite{booijInfluenceSuperimposedSteady1966a}.

\section{Conclusions}

We have shown that, within a constitutive model recently derived from microscopic dynamics, many aspects of the steady and transient rheology of concentrated soft suspensions (the von Mises yield criterion, the Trouton ratio, viscoelasticity in flow and transients under steps in shear rate or shear stress) depend strongly on the interplay between the elastic relaxation of the microstructure and advection. 
In particular the vorticity leads to significant misalignement of the microstructure (characterized by a fabric tensor) with respect to the strain rate tensor.
This in turn leads to a strong deviation from the quasi-Newtonian behavior often assumed in the literature~\cite{basterfieldInterpretationOrificeExtrusion2005,ovarlezThreedimensionalJammingFlows2010,shaukatShearMediatedElongational2012,coussotYieldStressFluid2014,zhangYieldingFlowSoftJammed2018}.
This calls for renewed efforts to measure experimentally  normal stress differences, in particular the first normal stress difference (the one affected by the vorticity) in simple shear. 
Only few pioneering works address this issue, but a consensus remains to emerge, the first normal stress difference  being either found positive and much smaller than the shear stress~\cite{habibiNormalStressMeasurement2016,de_cagny_yield_2019} or on the contrary negative and as large if not larger in amplitude than the shear stress~\cite{thompsonYieldStressTensor2018,thompsonRheologicalMaterialFunctions2020}. 

If such consensus does not exist, in the sense that normal stresses are found very much material dependent, or if the consensus is that actually normal stresses are small under simple shear, 
it would imply that the vorticity, which appears in a prescribed, non-negotiable manner in tensorial stress evolution because of frame indifference, is not the dominant contributor to the orientation of the fabric and/or stress.
This situation is also seen for suspensions of non-Brownian hard particles below their jamming point, where the first normal stress difference is known to be small compared to the shear stress~\cite{dennRheologyNonBrownianSuspensions2014}. 
Frame-indifferent stress evolution models unsurprisingly tend to overestimate it~\cite{goddardDissipativeAnisotropicFluid2006,chackoShearReversalDense2018,gillissenConstitutiveModelTimeDependent2019}, 
though this is not always the case, as some phenomenological models give good predictions, with careful adjustments of key parameters~\cite{phan-thienConstitutiveEquationConcentrated1995,phan-thienNewConstitutiveModel1999,ozendaTensorialRheologicalModel2020}.
Looking at how the predictions depend on said parameter values, we see that both Phan-Thien~\cite{phan-thienConstitutiveEquationConcentrated1995,phan-thienNewConstitutiveModel1999} and Ozenda et al.~\cite{ozendaTensorialRheologicalModel2020} find a large sensitivity of the first normal stress difference on the parameter governing the strength of the stress or fabric relaxation, corresponding to $[\beta(\phi)-\xi(\phi)\bm{\Sigma}':\bm{\Sigma}']$ in \eqref{eq:Sigmaprim}.
This is consistent with the idea that elastic relaxation must remain dominant over the rotation due to the vorticity for the first normal stress difference to remain small. 
This points to possible guidelines for improving the closures involved in the derivation of our model, if needed.

Interestingly, concentrated Brownian suspensions below jamming show a negative value of $N_1$ in simple shear, associated with a microstructure anisotropy misaligned from the principal axis of the strain-rate tensor, only in a direction opposite to the vorticity~\cite{fossStructureDiffusionRheology2000,morrisMicrostructureSimulatedBrownian2002,nazockdast_pair-particle_2013}. 
A careful analysis of the origin of this counter-rotation reveals that the microstructure is bimodal, with two directions for near-interactions accumulations. 
The first one lies between the compressional and flow gradient axes (as expected from a rotation by the vorticity), but the second one is along the flow direction, and corresponds to the increased probability of having particles following each other along streamlines~\cite{morrisMicrostructureSimulatedBrownian2002, nazockdast_pair-particle_2013}. 
In concentrated systems under simple shear, this flow ordering takes over the statistics of near interactions and yields the apparent counter-rotation of the second moment of the microstructure.
An approach based on a fabric tensor such as ours cannot capture a bimodal microstructure, which requires at least a fourth-ranked tensor descriptor~\cite{chackoShearReversalDense2018}.
Nonetheless, it is unclear at this stage if a similar phenomenon occurs for soft jammed suspensions. 
A contact accumulation along the flow direction may be observed in a jammed microgel suspension, especially during flow cessation transients~\cite{mohan_microscopic_2013}, but $N_1$ is found positive in the same system~\cite{liu_universality_2018}, indicating that flow-aligned contacts may not play a crucial role in the rheology, perhaps because they carry lower forces than contacts closer to the compression or gradient directions.

At a more fundamental level, it is remarkable that our constitutive model, which is based on assumptions of well-developed, homogeneous flow, can capture at least qualitatively 
many nontrivial aspects of the rheology of soft jammed suspensions usually attributed to 
possibly more complex or at least conceptually quite different microscopic mechanisms such as localized plastic events or spatial cooperativity. 
Of course it is entirely possible that these mechanisms, once coarse-grained,  give rise to a stress evolution equation structurally close to \eqref{eq:Sigmaprim} (and indeed Hand theory~\cite{hand_theory_1962} would argue for any constitutive model to be generically resembling  \eqref{eq:Sigmaprim}).
It would then not be a surprise that such equation would give qualitatively similar predictions to ours, and to some extent this would prevent the possibility to decide what is the right microscopic picture from macroscopic measurements.
This calls for more precise measurements of microstructural aspects of soft suspensions under flow, in experiments and numerical simulations, in order to guide further theoretical developments.

\section*{Conflicts of interest}
There are no conflicts to declare.

\section*{Acknowledgements}
This work is supported by the French National Research Agency in the framework of the "Investissements d'avenir" program (ANR-15-IDEX-02).
We thank Morton Denn for pointing out to us the polymer solution literature on negative storage modulus in parallel superposition.

\section*{Appendix~A: main steps of the derivation of the stress evolution equation}

We sketch in this Appendix the key steps of the derivation of the stress evolution equation, \eqref{eq:Sigmaprim}.
More details can be found in Ref.~\cite{CunyPRL21,CunyPRE21}.
The starting point is the exact evolution equation for the pair correlation function $g(\R{})$, which reads as
\begin{equation}
\begin{split}
\label{eq_g2}
\partial_t g(\R{}) + \bm{\nabla} \gdot \Big[ & \left(\DUinf \gdot \R{}\right) g(\R{}) -\gF(\R{})g(\R{})  \\ 
&-\rho \int \gF(\bm{r}') g_3(\R{},\bm{r}') \D \bm{r}'\Big] = 0 \,,
\end{split}    
\end{equation}
\eqref{eq_g2} is not closed in terms of $g(\R{})$, but also involves the three-body correlation function
$g_3(\bm{r},\bm{r}')$.
Defining the particle stress tensor $\gSigma$ from the virial formula \cite{nicot_definition_2013} as
\begin{equation}
	\label{sigma_int}
		\gSigma = \frac{\rho^2}{2} \int \big(\R{} \otimes \gF(\R{}) \big) \, g(\R{}) \, \D\R{},
\end{equation}
one can obtain an evolution equation for the stress tensor by multiplying \eqref{eq_g2} by $\frac{1}{2}\rho^2 \, \R{} \otimes \gF(\R{})$ and integrating over $\R{}$.
One then finds
\begin{equation}
	\label{eq:Sigma:v0}
	\dot{\gSigma} =\nabla \op{u}{\infty}{} \gdot \gSigma + \gSigma \gdot \nabla \op{u}{\infty\,T}{} + \bm{H}_2 - \bm{H}_3
\end{equation}
where the tensors $\bm{H}_2$ and $\bm{H}_3$ are defined by integrals over the pair and three-body correlation functions respectively,
\begin{align}
	\label{def_G2}
	\bm{H}_2 &= \frac{\rho^2}{2} \int \left[ \big(\Einf:\eroer\big)
	\left( (\R{}\otimes\R{})\cdot\nabla \gF(\R{})-\R{}\otimes\gF(\R{})\right)\right.\nonumber\\
		& \qquad \left. -\gF(\R{})\otimes\gF(\R{}) -\big(\R{}\otimes\gF(\R{})\big)\cdot\nabla \gF(\R{})^{T} \right] g(\R{}) \D\R{}\,, \\
	\label{def_G3}
	\bm{H}_3 &=\frac{\rho^3}{2} \iint \left[\gF(\bm{r}')\otimes \gF(\R{}) +
	\left(\R{} \otimes \gF(\bm{r}')\right)\gdot \nabla\gF(\R{})^{T}\right] \nonumber \\ & \qquad \qquad \qquad \qquad \qquad \qquad \quad \times g_3(\R{},\bm{r}')\D\R{}\D\bm{r}'\,.
\end{align}
A corresponding evolution equation for the deviatoric part $\gSigma'$ of the stress tensor is obtained by taking the deviatoric part of \eqref{eq:Sigma:v0}.
However, the resulting equation is not closed in terms of $\gSigma'$.
The closed equation \eqref{eq:Sigmaprim} is then obtained after a set of approximations, as detailed in Ref.~\cite{CunyPRL21,CunyPRE21}.
(i) The three-body correlation function $g_3(\bm{r},\bm{r}')$ is expressed in terms of the pair correlation function $g(\bm{r})$ through the Kirkwood approximation,
\begin{equation}
g_3(\bm{r},\bm{r}') \approx g(\bm{r}) g(\bm{r}') g(\bm{r}-\bm{r}')\, .
\end{equation}
(Other choices are possible, see for instance~\cite{nazockdast_microstructural_2012,jenkinsPredictionsMicrostructureStress2021}.)
(ii) The anisotropic part of the pair correlation function is expanded to lowest order, that is to the second harmonic.
(iii) The isotropic part of the pair correlation is given a schematic form with an infinitely thin peak for the first neighbor shell, and a uniform sea of particles for larger distances; this simplified form can be parametrized only in terms of the pressure.
(iv) The pressure, whose evolution equation is obtained by taking the trace of \eqref{eq:Sigma:v0}, turns out to be a fast dynamical variable and can be eliminated using an equation of state $p(\phi)$, where $\phi$ is the surface fraction of particles.

\section*{Appendix~B: strain rate for generic 2D flow}

We here briefly show the origin of our scalar parametrization~\eqref{eq:flow_family}.
In the main text, we presented in~\eqref{eq:flow_family} a scalar parametrization of a steady uniform 2D flow.
For a generic 2D flow (possibly unsteady and non uniform), a generalization of this paramatrization is useful only if it is frame-indifferent, that is, the scalar characterizing the flow is invariant under a time dependent rotation of the flow~\cite{schunkConstitutiveEquationModeling1990,giusteriTheoreticalFrameworkSteadystate2018}. 
Generically, a flow is characterized by its local velocity gradient $\nabla \bm{u}^\infty$. 
The velocity gradient itself is not frame-indifferent, because its antisymmetric part, the vorticity, is not.
One should thus not parametrize the velocity gradient, but only its frame-indifferent bit, defined as follows~\cite{schunkConstitutiveEquationModeling1990,giusteriTheoreticalFrameworkSteadystate2018}
\begin{equation}
 \left(\nabla \bm{u}^\infty\right)^\prime = \nabla \bm{u}^\infty - \bm{\Omega}^\infty_\mathrm{SR}\, ,
\end{equation}
where $\bm{\Omega}^\infty_\mathrm{SR}$ is the local spin tensor of the eigenvectors of $\bm{E}^\infty$, that is, the part of the total spin $\bm{\Omega}^\infty$ which corresponds to a local solid rotation. 
Said otherwise, the frame indifferent part of the local velocity gradient is the velocity gradient in a  frame in which the local strain rate tensor does not rotate.
The spin tensor $\bm{\Omega}^\infty_\mathrm{SR}$ is defined as $\bm{\Omega}^\infty_\mathrm{SR}\cdot \bm{x} = \bm{\omega}\times \bm{x}$ for any vector $\bm{x}$, with~\cite{schunkConstitutiveEquationModeling1990} 
\begin{equation}
    \bm{\omega} =\bm{e} \times \left(\frac{\partial \bm{e}}{\partial t} + \bm{u}^\infty \cdot \nabla \bm{e}\right),
\end{equation}
with $\bm{e} = \bm{e}_1, \bm{e}_2$, one of the unit eigenvectors of $\bm{E}^\infty$ (these eigenvectors are orthogonal, such that they share the same spin.) 
From this we can define a frame-indifferent flow  shape $\bm{K}^\infty = (\nabla \bm{u}^\infty)^\prime/\sqrt{(\nabla \bm{u}^\infty)^\prime:(\nabla \bm{u}^\infty)^\prime}$. 
(For the special case of steady uniform flows considered in the main text $\left(\nabla \bm{u}^\infty\right)^\prime = \nabla \bm{u}^\infty$, and thus $\bm{K}^\infty = \nabla \bm{u}^\infty/\sqrt{\nabla \bm{u}^\infty:\nabla \bm{u}^\infty}$.)

The set of possible 2D flows of course include planar extension and simple shear.
We look for parametrization following the usual conventions for the principal axes of these two specific flows, corresponding for planar extension to
\begin{equation}
    \bm{K}^\infty = 
                \begin{pmatrix}
                    1/\sqrt{2} &  0 \\
                    0 &  -1/\sqrt{2}
    \end{pmatrix}\, ,
    \label{eq:K_extension}
\end{equation}
and for simple shear to
\begin{equation}
    \bm{K}^\infty = 
                \begin{pmatrix}
                   0 &  1 \\
                    0 &  0
    \end{pmatrix}\, .
    \label{eq:K_simple_shear}
\end{equation}

We can generically express $\bm{K}^\infty$ in terms of the eigenvectors of $\bm{E}^\infty$,
\begin{equation}
    \bm{K}^\infty = a\bm{e}_1 \otimes \bm{e}_1 + b\bm{e}_2 \otimes \bm{e}_2
    + c \bm{e}_1 \otimes \bm{e}_2 + d \bm{e}_2 \otimes \bm{e}_1\, ,
\end{equation}
with $a$, $b$, $c$ and $d$ four scalars.
Now, because of incompressibility $\bm{K}^\infty$ is traceless, which imposes $a=-b$.
Furthermore, $\bm{e}_1$ and $\bm{e}_2$ are eigenvectors of $\bm{E}^\infty$, which implies that they must also be eigenvectors of the symmetric part of $\bm{K}^\infty$.
A quick calculation concludes that this imposes $c=-d$.
Finally, $\bm{K}^\infty:\bm{K}^\infty = 1$, which implies that we can parametrize
$\bm{K}^\infty$ with an angle $\alpha$ as
\begin{equation}
\bm{K}^\infty = \frac{1}{\sqrt{2}}\left[\cos\alpha (\bm{e}_1 \otimes \bm{e}_1 - \bm{e}_2 \otimes \bm{e}_2) + \sin\alpha(\bm{e}_1 \otimes \bm{e}_2 - \bm{e}_2 \otimes \bm{e}_1)\right]\,  .
    \label{eq:K_E_basis}
\end{equation}

By identification, planar extension corresponds to $\alpha = 0$, and simple shear to $\alpha = \pi/4$. 
The simplest family of flow shapes that interpolates between the convention \eqref{eq:K_extension} for planar extension in $\alpha=0$ and the convention \eqref{eq:K_simple_shear} for simple shear in $\alpha = \pi/4$ is a family in which the eigenvectors of $\bm{E}^\infty$ are rotated by $\alpha$ with respect to the $x$-$y$ axes.
In the $x$-$y$ basis, this family corresponds \eqref{eq:flow_family} in the main text.





\providecommand*{\mcitethebibliography}{\thebibliography}
\csname @ifundefined\endcsname{endmcitethebibliography}
{\let\endmcitethebibliography\endthebibliography}{}

\end{document}